\documentclass[final]{siamltex}

\renewcommand{\small}{}

\usepackage[section]{algorithm}
\usepackage[noend]{algorithmic}
        \newcommand{\INPUT}[1]{\item[\algorithmicinput]{\em #1}}
        \newcommand{\OUTPUT}[1]{\item[\algorithmicoutput]{\em #1} \smallskip\par}
        \newcommand{\algorithmicinput}{\textbf{\small ~~~given:}}
        \newcommand{\algorithmicoutput}{\textbf{\small ~return:}}

\newenvironment{subroutine}{\floatname{algorithm}{Subroutine}\begin{algorithm}}{\end{algorithm}}

\newfloat{subsubroutine}{htbp}{loa}[algorithm]
\floatname{subsubroutine}{Subroutine}

\newfloat{PTAS}{htbp}{loa}
\floatname{PTAS}{PTAS}

\makeatletter
\newcommand{\currentalgorithmname}{\fname@algorithm}
\makeatother

\usepackage{graphicx}
\usepackage{xspace}
\usepackage{wrapfig}
\usepackage{amssymb} 

\newcommand{\operatorname}[1]{\mathop{\mathrm{#1}}}
\newcommand{\text}[1]{{\ensuremath{\textrm{{#1}}}}}

\renewcommand{\endproof}{\hfill \vbox{\hrule height0.6pt\hbox{%
   \vrule height1.3ex width0.6pt\hskip0.8ex
   \vrule width0.6pt}\hrule height0.6pt
  }}

\newtheorem{subdefinition}[theorem]{Definition}
\newtheorem{sublemma}[theorem]{Lemma}
\newtheorem{subobservation}[theorem]{Observation}

\newcommand{\calI}{{\cal I}}

\newcommand{\calL}{{\cal L}}

\newcommand{\calR}{{\cal R}}

\newcommand{\calU}{{\cal U}}

\newcommand{\calX}{{\cal X}}
\newcommand{\calY}{{\cal Y}}

\newcommand{\R}{\mathbb{R}}
\newcommand{\Rp}{\R_{^{\ge 0}}}
\newcommand{\N}{{\mathbb{N}}}

\newcommand{\Nz}{\N_{^{\ge 0}}}

\newcommand\eps\epsilon
\renewcommand\varepsilon\eps

\newcommand{\margin}[1]{{\marginpar{\parbox{0.65in}{\raggedright\footnotesize #1\par}}}}
\newcommand{\notedefn}[1]{{\margin{{$\triangleright$ #1}}}}
\renewcommand{\notedefn}[1]{}

\newcommand{\envtitle}{}

\newcommand{\proc}[1]{\text{\small\sf #1}}

\newcommand{\lift}[1]{\overline{#1}}
\renewcommand{\lift}[1]{\hat{#1}}
\renewcommand{\lift}[1]{{#1^{{\!}^{_\uparrow}}}}
\renewcommand{\lift}[1]{{#1}'}

\newcommand{\enc}[1]{\operatorname{enc}({#1})}

\newcommand{\cost}{\operatorname{cost}}

\newcommand{\prefixes}{\operatorname{prefixes}}

\newcommand{\polylog}{\operatorname{polylog}}
\newcommand{\poly}{\operatorname{poly}}

\newcommand{\pad}{\operatorname{pad}}
\newcommand{\chunk}{\operatorname{chunk}}
\newcommand{\unchunk}{\operatorname{unchunk}}
\newcommand{\minimize}{\operatorname{minimize}}

\newcommand{\problem}[1]{{\sc #1}\xspace}
\newcommand{\opt}{\ensuremath{\operatorname{\mbox{\sc opt}}}\xspace}
\newcommand{\smallopt}{{\ensuremath{\mbox{\sc\tiny opt}}}}
\renewcommand{\smallopt}{{\ensuremath{\star}}}

\newcommand{\group}{G}
\newcommand{\relaxedCode}{\calA}
\newcommand{\reservedCode}{\calB}
\newcommand{\code}{\calX}

\newcommand{\instance}{\calI}
\newcommand{\limitedlevels}{\calC}
\renewcommand{\limitedlevels}{\calL}

\newcommand{\universe}{\calU}
\newcommand{\roots}{\calR}

\newcommand{\letter}{\ell}
\newcommand{\tree}{T}
\newcommand{\forest}{F}
\newcommand{\prob}{p}
\newcommand{\PROB}{P}
\newcommand{\alphabet}{\Sigma}

\newcommand{\threshold}{\tau}

\newcommand{\HUL}{\problem{Hulc}}
\newcommand{\HULR}{$\threshold$-\HUL}
\renewcommand{\HULR}{$\threshold$-\problem{Relax}}

\newcommand{\MP}{\text{\sc Karp}}
\newcommand{\MPG}{\text{\sc ilp}}

\newcommand{\correctness}{\smallskip\noindent{\bf Correctness.}\xspace}
\renewcommand{\time}{\smallskip\noindent{\bf Time.}\xspace}

\newenvironment{LP}[1]
{\(\begin{array}[t]{r@{~~}c@{~~}ll}
\multicolumn{4}{r}{ \text{#1}}
\\[-2ex]
\multicolumn{4}{@{}l}{\minimize~ \sum_{i,k} \,\prob_k\, i\,y_{ki} \text{~~s.t.} \hphantom{~~~\text{#1}}}
\\[1.1ex]}
{\\[0.5ex]\end{array}\)}

\newenvironment{LPNoCost}[1]
{\(\begin{array}[t]{r@{~~}c@{~~}ll}
\multicolumn{4}{@{~}l}{ \text{#1}}\\[3pt]}
{\\[0.5ex]\end{array}\)}

\newenvironment{LPTAB}
{\begin{center}\begin{tabular}{@{}|l@{}|@{}l@{}|@{}} \hline}
{\\ \hline \end{tabular}\end{center}}

\begin{document}

\title{Huffman Coding with Letter Costs:
\\A Linear-Time Approximation Scheme
\thanks{Full version in SICOMP.  Conference version appeared as
  ``Huffman Coding with Unequal Letter Costs''
  in STOC'02.}}

\author{
Mordecai J. Golin
\thanks{
Partially supported by HK RGC Competitive
Research Grants HKUST 6137/98E, 6162/00E and 6082/01E;
        {Hong Kong UST},
        {Clear Water Bay},
        {Kowloon, Hong Kong},
        {\tt golin@cs.ust.hk}.
}
\and
Claire Mathieu
\thanks{
  Brown University; Providence, Rhode Island, USA.
  Partially supported by NSF grant CCF-0728816.
}
\and
Neal E. Young
\thanks{
University of California, Riverside; Riverside, California, USA.
Partially supported by NSF grants  CNS-0626912, CCF-0729071.
}
}
\maketitle

\begin{abstract}
We give a polynomial-time approximation scheme
for the generalization of Huffman Coding
in which codeword letters have non-uniform costs
(as in Morse code, where the dash is twice as long as the dot).
The algorithm computes a
$(1+\eps)$-approximate solution
in time $O(n + f(\eps) \log^3 n)$,
where $n$ is the input size.
\end{abstract}



\pagestyle{myheadings}
\thispagestyle{plain}
\markboth{GOLIN, MATHIEU, AND YOUNG}{HUFFMAN CODING WITH LETTER COSTS}

\section{Introduction}\label{sec:intro}


The problem of constructing a minimum-cost prefix-free code for a given distribution, 
known as { Huffman Coding}, is well-known and admits a simple greedy algorithm.  
But there are many well-studied variations of this simple problem for which fast algorithms are not known.
This paper considers one such variant
--- the generalization of { Huffman Coding} in which the encoding letters have non-uniform costs --- 
for which it describes a polynomial-time approximation scheme (PTAS).

Letter costs arise in coding problems where different characters have
different transmission times or storage costs
\cite{Blach-54, Marcus-57, Karp-61, Stan-70, Varn-71}.
One historical example is the {telegraph channel} --- Morse code. 
There, the encoding alphabet is 
$\{ \cdot , - \}$ and dashes are twice as long as dots, i.e. 
$\cost(-)=2\cost(\cdot)$
\cite{gilb-69, gilb-95,Krause-62}.
A simple data-storage example is the $(h,k)$-run-length-limited codes used in magnetic and optical storage.
There, the codewords are binary and constrained
so that each `1' must be preceded by at least $h$, and at most
$k$, `0's
\cite{Imm-99,GR-98}.
(To reduce this problem to Huffman Coding with letter costs,
use an encoding alphabet with one letter  of cost $j+1$ for each string `$0^{j} 1$',  where $h\leq j\leq k$.)

\begin{figure*}[t]

\centering
\includegraphics[height=20ex]{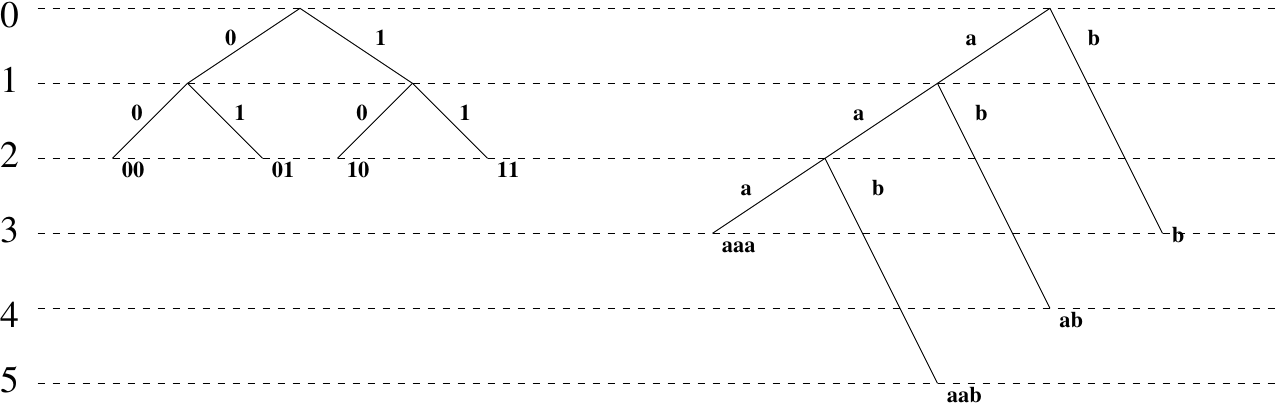}
\caption{\em Two prefix-free codes, in tree representation.
  The letter costs are $(1,1)$ and $(1,3)$, respectively. 
The code on the left is
$\{00,01,10,11\}$.
The code on the right is
$\{aaa,b,ab,aab\}$.
The costs of the two codes are, respectively
$2(\prob_1+\prob_2+\prob_3+\prob_4)$
and $3(\prob_1+\prob_2) + 4 \prob_3 +5p_4.$
}\label{fig:example}
\hrulefill
\end{figure*}

\begin{definition}[\envtitle Huffman Coding with Letter Costs -- \HUL]
The input is
\begin{itemize}
\item a probability distribution $\prob$ on $[n]$,
\item a {\em codeword alphabet} $\alphabet$ of size at most $n$,
\item  for each letter $\letter\in\alphabet$, a specified non-negative integer\footnote
{The assumption of integer costs is for technical reasons.  
In fact the algorithm given here handles arbitrary real letter costs.
See Section~\ref{subsec:coarse}.}
$\cost(\letter)$. 
\end{itemize}
\smallskip

The output is a code $\code$ consisting of $n$ codewords, where  $\code_i\in\alphabet^*$ is the codeword for probability $\prob_i$. The code must be prefix-free.   (That is, no codeword is a prefix of any other.)
The goal is to minimize the cost of $\code$, which is denoted $\cost(\code)$ and defined to be
 $\sum_{i=1}^n \prob_i \,\cost(\code_i)$, where, for any string $w$, $\cost(w) $ is the sum of the costs of the letters in $w$.
(See Fig.~\ref{fig:example}.)
\end{definition}\smallskip

\notedefn{code}
\notedefn{prefix-free}


 \HUL has been extensively studied.
Blachman \cite[(1954)]{Blach-54}, Marcus \cite[(1957)]{Marcus-57},
and Gilbert \cite[(1995)]{gilb-95} give heuristic algorithms.  
The first algorithm yielding an exact solution is due to Karp,
based on integer linear programming \cite[(1961)]{Karp-61}.
Karp's algorithm does not run in polynomial time.
A number of other works use some form of entropy
to lower bound the optimal cost \opt,
and give polynomial-time algorithms that compute heuristic solutions
of cost at most $\opt + f(\cost)$
where $f(\cost)$ is some function of the letter costs
\cite[(1962-2008)]{Krause-62, Csiszar-69, Cot-77, Mehlhorn-80, AM-80, golin2008more}.
These algorithms are not constant-factor approximation algorithms,
even for fixed letter costs,
because non-trivial instances can have small \opt.
For further references and other uses of \HUL, 
see Abrahams'  survey on source coding
\cite[Section 2.7]{Abrahams-01}.

However, there is no known polynomial-time algorithm for \HUL,
nor is it known to be NP-hard.
Before now, the problem was not known to
have any polynomial-time constant-factor approximation algorithm.
Our main result is a polynomial-time approximation scheme:

\begin{theorem}[PTAS for \HUL]\label{thm:main}
Given any \HUL instance,
the tree representation of a prefix-free code of cost at most $1+O(\eps)$ times minimum
can be computed in time $O(n) + O_\eps(\log^3 n)$.
\end{theorem}\smallskip

The tree representation is a standard representation of prefix-free codes (see Defn.~\ref{def:tree}
and Fig.~\ref{fig:example}).
In the $O_\eps(\log^3 n)$ term,
the subscript $\eps$ denotes
that the hidden constant in the big-O depends on $\eps$.

We note without proof that the above PTAS can easily be adapted to
show that, given any fixed $\eps$, the problem of $(1+\eps)$-approximating \HUL 
is in NC (Nick's class --- polynomially many parallel processors
and polylogarithmic time).

\paragraph{Related problems}

When all letter costs are equal, \HUL reduces to standard {\sc Huffman Coding}.
The well-known greedy algorithm for {\sc Huffman Coding} is due to Huffman  \cite{Huff-52}.
The algorithm runs in $O(n)$ time, or $O(n\log n)$ time if $\prob$ is not sorted.

When the letter costs are fixed integers, 
Golin and Rote give a
dynamic programming algorithm that produces exact solutions in
time $O(n^{2+\max_j \cost(\letter_j)})$ \cite{GR-98}.  This is improved to
$O(n^{\max_j \cost(\letter_j)})$ for alphabets of size 2 by Bradford et al.\ \cite{BGLR-02}
and for general (but fixed) alphabets by Dumitrescu \cite{dumitrescu2006faster}.

When all the {\em probabilities} are equal
(each $\prob_j = 1/n$), \HUL is the 
 \problem{Varn Coding} problem, which is solvable in polynomial time
\cite{Varn-71, LEC-73, Cot-75, PGE-75, GY-94, CG-01}.

Finally, \problem{Alphabetic Coding} is like \problem{Huffman Coding}
but with an additional constraint on the code: the order of the given probabilities matters
---
their respective codewords must be in increasing alphabetic order.
(Here the probabilities are not assumed to be in sorted order.)
\problem{Alphabetic Coding with Letter Costs}
(also called \problem{Dichotomous Search} \cite{Hind-90}
or the \problem{Leaky Shower} problem \cite{kr-89})
models designing testing procedures
where the time required by each test depends upon the outcome
\cite[(\S 6.2.2; ex.\ 33)]{Knuth-73}.
That problem has a polynomial-time algorithm \cite{Itai-76}.


\paragraph{Basic idea of the PTAS}
To give some intuition for the PTAS, consider the following simple idea.
Without the prefix-free constraint, \HUL would be easy to solve: 
to find an optimal code $\code$, one could simply enumerate the strings in $\alphabet^*$ 
in order of increasing cost, and take $\code_i$ to be the $i$th string enumerated.

The cost of this optimal non-prefix-free code
$\code$ is certainly a lower bound on the minimum-cost of any prefix-free code.
Now consider modifying $\code$ to make it prefix-free as follows.  
Prepend to each codeword $\code_i$
its length, encoded in a prefix-free binary encoding.  That is, take 
$\code'_i = \enc{|\code_i|}\,\code_i$,  where $\enc \ell$ 
is any natural prefix-free encoding of integer $\ell$.
(For example, make the standard binary encoding prefix-free 
by replacing `0' and `1' by `01' and `10', respectively,
then append a `00'.)  The resulting code is prefix-free, 
because knowing the length of an upcoming codeword
is enough to determine where it ends.  And, intuitively,
the cost of $\code'$ should not exceed the cost of $\code$ by much, because each codeword
in $\code$ with $\ell$ letters only has $O(\log_2\ell)  \le O(\eps \ell)$ letters added to it.
Thus, the cost of prefix-free code $\code'$ should be at most $1+O(\eps)$
times the cost of $\code$, and thus at most $1+O(\eps)$ times the cost of $\opt$.

Why does the above idea fail?  
It fails because $\log_2\ell$ is not $O(\eps \ell)$ when $\ell< O(\eps^{-1}\log \eps^{-1})$.
That is, when a codeword is small, prepending its length can increase its cost by too much.
To work around this, we handle the small codewords separately,
determining their placement by exhaustive search.  
This is the basic idea of the PTAS.  
The rest of the paper gives the technical details.

\subsection*{Terminology and definitions}
\label{sec:prelim}

For technical reasons, we work with a generalization of \HUL
in which codewords can be restricted to a given universe $\universe$:
\begin{definition}[\HUL with restricted universe]\label{def:roots}
The input is a \HUL instance $(\prob,\alphabet,\cost)$ and a {\em codeword universe} $\universe\subseteq\alphabet^*$.
The universe $\universe$ is specified by a finite, prefix-free set $\roots\subset \alphabet^*$ of ``roots''
such that $\universe$ consists of  the strings with a prefix in $\roots$.
 The problem is to find a code of minimum cost among the prefix-free codes 
whose codewords are in $\universe$.
\end{definition}\smallskip

Formally, $\universe$ is defined from the given root set $\roots\subset\alphabet^*$ as
the set of strings $x\in\alphabet^*$ such that 
$\prefixes(x) \cap \roots\ne \emptyset$,
where $\prefixes(x)$ denotes the set of all prefixes of $x$. 
The universe is necessarily closed under appending letters
(that is, if $x\in\universe$ and $y$ has  $x$ as a prefix, then $y\in\universe$).
If $\universe = \alphabet^*$ (i.e., $\roots$ contains just the empty string), 
then the problem is \HUL\ as defined at the start of the paper.


In any problem instance, we assume the following without loss of generality:
\begin{itemize}
\item There are at most $n$ letters in the alphabet $\alphabet$, and they are $\{0,1,\ldots,|\alphabet|-1\}$.
\item The letter costs are increasing:
$\cost(0) \le \cost(1) \le \cdots \le \cost(|\alphabet|-1)$.
\\(If not, sort them first, adding $O(n\log n)$ or less to the run time.)
\item The codeword probabilities are decreasing: $\prob_1 \ge \prob_2 \ge \cdots \ge \prob_n$.
\\(If not, sort them first, adding $O(n\log n)$ to the run time.)
\end{itemize}
\smallskip

\begin{definition}[\envtitle monotone code]
A code $\code$ is {\em monotone} if 
\[\cost(\code_1) \le \cost(\code_2) \le \cdots \le \cost(\code_n).\]
\end{definition}
\noindent
For any code $\code$, reordering its codewords to make it monotone 
does not increase its cost (since $\prob$ is decreasing),
so we generally focus on monotone codes.

\medskip

Next we define two more compact representations of codes:

\begin{definition}[\envtitle signature representation]
Given a set $\code\subseteq \alphabet^*$,
its  {\em signature} is the vector $x$ such that $x_i$ 
is the number of strings in $\code$ that have cost $i$.
(Recall that letters, and thus codewords, have integer costs.)
\end{definition}\smallskip

In Fig.~\ref{fig:example}, the first code has signature $(0,0,4)$;
the second code has signature $(0,0,0,2,1,1)$.

Many codes may have the same signature, but any two (monotone)
codes with the same signature are essentially equivalent.
For example,
the signature $x$ of a monotone code $\code$  determines $\cost(\code)$: 
indeed, $\cost(\code_k)=i(k)$
where $i(k)$ is the minimum $i$ such that $x_1+\cdots + x_{i} \ge k$.



\begin{definition}[tree representation]\label{def:tree}
The {\em tree representation} of a code $\code$ is a forest 
with a node $v(s)$ for each string $s\in \prefixes(\code)\cap\universe$,
and an edge from each (parent) node $v(s)$ to (child) node $v(s')$
if $s'=s\ell$ for some letter $\ell\in\alphabet$.
Each root of the forest is labeled with its corresponding string in $\roots$.
\end{definition}\smallskip

For standard Huffman coding (with just two equal-cost letters $\{0,1\}$
and $\universe = \alphabet^*$), the tree representation is a binary tree.
Each codeword traces a path from the root, 
with `$0$'s corresponding to left edges and `$1$'s  to right edges.   
See, for example, $\code_1$ in Fig.~\ref{fig:example}.
If $\universe \ne \alphabet^*$, the tree representation can be a forest
(that is, it can have multiple trees, each with a distinct root in $\roots$).

A code is prefix-free if and only if,
in its tree representation, all codewords are leaf nodes.

\begin{definition}[levels]
The {\em $i$th level} of a set $\code\subseteq \alphabet^*$ contains the cost-$i$ strings in $\code$.
(See the horizontal lines in Fig.~\ref{fig:example}.)
\notedefn{levels}
\end{definition}\smallskip

\paragraph{Additional terminology and notation}
Throughout the paper $\eps$ is an arbitrary constant strictly between 0 and $1/2$.
The PTAS returns a near-optimal code
--- a code of cost $1+O(\eps)$ times the minimum cost of any prefix-free code.
The terms ``nearly'', ``approximately'', etc.\ 
generally mean ``within a $1+O(\eps)$ factor''.
The notation $O_\eps(f(n))$ denotes $O(f(n))$, 
where the hidden constant in the big-O can depend on $\eps$.

Given a problem instance $\instance$,
the cost of an optimal solution is denoted $\opt(\instance)$,
or just $\opt$ if $\instance$ is clear from context.
As is standard, $[n]$ denotes $\{1,2,\ldots,n\}$.
We let $[i..j]$ denote $\{i,i+1,\ldots,j\}$.

\begin{figure}[t]
\framebox{\parbox{0.98\textwidth}{
\setlength{\parskip}{0.9ex}
\renewcommand{\baselinestretch}{0.94}
\small
\caption{{\bf Outline of the proof of Thm.~\ref{thm:main} (PTAS for \HUL)}}\label{fig:outline}

{\bf Section~\ref{sec:cost1sig}.}~~For instances in which $\cost(1)\le 3/\eps$,
the signature $x$ of a near-optimal prefix-free code can be computed
in time $O_\eps(\log^2 n)$,
provided the following inputs are precomputed: the cumulative probability distribution $\PROB$ 
(for the distribution $\prob$)
and the signatures $\sigma$ and $r$ of, respectively, the alphabet $\alphabet$
and the roots $\roots$ of the universe $\universe$.
(These inputs $\prob$, $\sigma$, and $r$ can be precomputed in $O(n)$ time.)

{\bf Section~\ref{sec:cost1tree}.}~~From the signature $x$,
the tree can be built in $O(n) + O_\eps(\log^2 n)$ time.

{\bf Section~\ref{sec:general}.}~~
Any arbitrary instance of \HUL reduces to $O_\eps(\log n)$ 
instances with $\cost(1)\le 3/\eps$, which can in turn be solved by the PTAS
from Sections~\ref{sec:cost1sig} and~\ref{sec:cost1tree},
giving the full PTAS.

\bigskip

\centerline{\bf Breakdown of Section~\ref{sec:cost1sig} (finding a near-optimal signature when $\cost(1)\le 3/\eps$)}

Sections~\ref{subsec:structural} --~\ref{subsec:MP}
define and analyze certain structural properties related to near-optimal codes.
Section~\ref{subsec:cost1sig proof} uses these properties to assemble
the PTAS for instances with $\cost(1)\le 3/\eps$.

\textbf{Section~\ref{subsec:structural}.} In a {\em $\threshold$-relaxed} code,
codewords of cost at least a given threshold $\threshold$
{\em are } allowed to be prefixes of other codewords.
For appropriate (constant) $\threshold$,
this relaxation 
(finding a min-cost $\threshold$-relaxed code)
has a gap of $1+O(\eps)$
--- a given $\threshold$-relaxed code can be efficiently ``rounded''
into a prefix-free code
without increasing the cost by more than a factor of $1+O(\eps)$.

Thus, it suffices to find
a near-optimal $\threshold$-relaxed code and then round it.

Any $\threshold$-relaxed code $\code$ is essentially determined by 
its set $\code_{<\threshold}$ codewords of cost less than $\threshold$.
This observation alone is enough to give a slow PTAS
for instances with $\cost(1)\le 3/\eps$:
exhaustively search the possible signatures $f$ of $\code_{<\threshold}$
to find the best.

This would give run time $n^{O_\eps(1)}$.
The remaining subsections improve the time
to $O(n) + O_\eps(\log^2 n)$.

\textbf{Section~\ref{subsec:grouping}.} Restricting attention to a relatively
small subset of $\threshold$-relaxed codes, so-called {\em group-respecting} codes,
increases the cost by at most a $1+O(\eps)$ factor.  Thus, it suffices to find
an optimal group-respecting $\threshold$-relaxed code.
This observation reduces the search space size to a constant.

\textbf{Section~\ref{subsec:limitedlevels}.} 
There is a logarithmic-size set $\limitedlevels$ of levels such that, 
without loss of generality, we can consider only codes with support in $\limitedlevels$ ---
that is, codes whose tree representations have (interior or codeword) nodes only
in levels in $\limitedlevels$.
Thus, it suffices to find an optimal group-respecting $\threshold$-relaxed code
with support in $\limitedlevels$.

\textbf{Section~\ref{subsec:MP}.}  
The problem of finding the {\em signature} of such a code
is formally modeled via integer linear program, $\MPG$.
Thanks to Section~\ref{subsec:limitedlevels}, $\MPG$ has logarithmic size.
Further, given the values of just a constant number of key variables of $\MPG$,
an optimal (greedy) assignment of the rest of the variables can easily be computed
in logarithmic time.

\textbf{Section~\ref{subsec:cost1sig proof}.} 
Putting the above pieces together,
the PTAS for instances with $\cost(1)\le 3/\eps$
enumerates the constantly many possible assignments of the key variables in $\MPG$,
then chooses the solution giving minimum cost.
This gives the signature $x$ of a near-optimal $\threshold$-relaxed code,
which is converted via the rounding procedure of Section~\ref{subsec:structural}
into the desired signature $x'$ of a near-optimal prefix-free code.
}}
\end{figure}

\smallskip
The rest of the paper proves Thm.~\ref{thm:main}.
The value of the second-largest letter cost, i.e., $\cost(1)$, 
is a major consideration in the proof.
We first describe a PTAS for the case when $\cost(1)\le 3/\eps$;
we then reduce the general case to that one.
For efficiency, the PTAS works mainly with code signatures;
in the last step, it converts the appropriate signature to a tree representation.

See Fig.~\ref{fig:outline} for a summary of the three remaining sections,
and the five subsections of Section~\ref{sec:cost1sig}.

\section{Computing the signature of a near-optimal code when $\cost(1)\le 3/\eps$}
\label{sec:cost1sig}

This section gives the core algorithm of the PTAS.
Given any instance in which $\cost(1) \le 3/\eps$,
the core algorithm computes the signature of a near-optimal prefix-free code for that instance.
(Recall that all letter costs are integers.)
Formally, in this section we prove the following theorem.
\newcommand{\propsig}{
Fix any instance $\instance=(\prob,\alphabet,\cost,\universe)$ of \HUL with restricted universe
such that $\cost(1) \le 3/\eps$.
Let $\PROB$ be
the cumulative probability distribution for $p$: $P_\ell = \sum_{k\le \ell} p_k$ (for $\ell\in[n]$).
Let $\sigma$ be the signature of $\alphabet$.
Let $r$ be the signature of the roots of $\universe$.
Assume that $\PROB$, $\sigma$, and $r$ are given as inputs.

Then
the signature and approximate cost of a  prefix-free code (for $\instance$) with cost
at most $(1 + O(\eps))\opt(I)$ 
can be computed in time $O_\eps(\log^2 n)$.
}
\begin{theorem}\label{thm:cost1sig}
\propsig
\end{theorem}\smallskip

\newcommand{\thmcostsig}{\par{\sc Theorem~\ref{thm:cost1sig}.} {\em \propsig} \par\smallskip}

Throughout this section, in proving Thm.~\ref{thm:cost1sig},
assume $\cost(1)\le 3/\eps$.
(The proof holds for any instance in which $\cost(1) = O_\eps(1)$;
we focus on the case $\cost(1) \le 3/\eps$
only because later we reduce the general case to that case.)


\subsection{Allowing codes to be $\tau$-relaxed}\label{subsec:structural}\label{subsec:relax}

\begin{figure*}[h]
\hrulefill

\centering
\caption{\em Tree representation of a $\threshold$-relaxed code $\code$
with four codewords in levels less than $\threshold=6$.
The 21 codewords in levels $\threshold$ and higher
can be prefixes of other codewords, so they are taken to
be the cheapest 21 strings that have no prefix in $\code$ of cost less than $\threshold$.
}
\label{fig:relax}

\includegraphics[height=20ex]{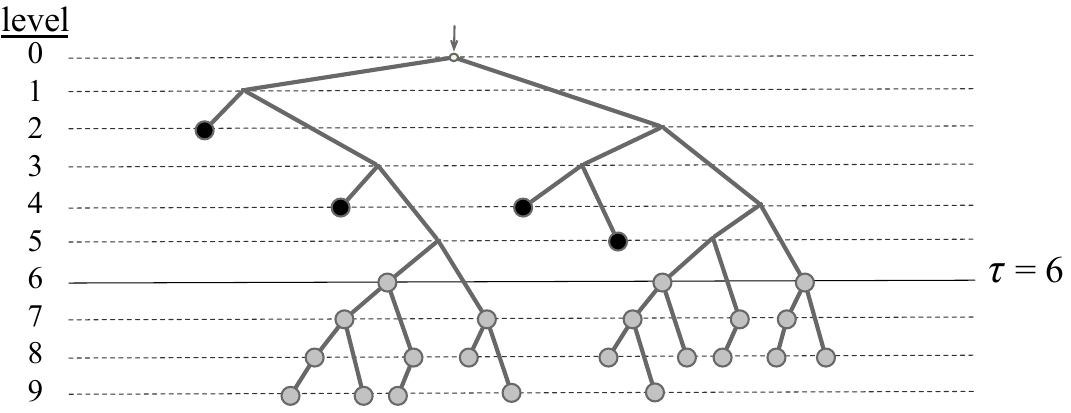}

\hrulefill
\end{figure*}

In a {\em $\threshold$-relaxed code},
codewords of cost at least $\threshold$
{\em can} be prefixes of other codewords
as illustrated in Fig.~\ref{fig:relax}.

\begin{definition}[relaxation \HULR]\label{def:hulr}
Given a {\em threshold} $\threshold\ge 0$, 
a code $\code$ is {\em $\threshold$-relaxed} 
if no codeword {\em of cost less than $\threshold$}
is the prefix of another codeword.
(Prefix-free codes are $\threshold$-relaxed,
but not vice versa.)

{\em \HULR} is the problem
of finding a minimum-cost $\threshold$-relaxed code
for a given instance of \HUL.
\end{definition}\smallskip

\HUL reduces to \HULR.
Specifically, 
if the threshold $\threshold$ is appropriately chosen,
the relaxation changes the optimal cost by at most a $1+O(\eps)$ factor:

\begin{lemma}[relaxation gap]\label{lemma:relax}
Fix threshold $\threshold=\lceil\log_2[\cost(1)/\eps] \cost(1)/\eps\rceil$.
Given a $\threshold$-relaxed code $\code$ for any \HUL instance,
there exists a prefix-free code $\code'$ such that $\cost(\code') = (1+O(\eps))\cost(\code)$.
The code $\code'$ is produced by calling procedure $\proc{Round}(\code)$.
\end{lemma}

\begin{proof}
The procedure \proc{Round} is Alg.~\ref{alg:relax}, below.
Roughly, for each codeword of cost $\threshold$ or more in $\code$,
\proc{Round} inserts the cost, $i$,
(encoded in a simple prefix-free binary code, as specified in Step~\ref{step:binary} of the algorithm)
into the codeword, starting at level $\threshold$.
For technical reasons, instead of the cost $i$, it actually inserts $i-\hat\threshold$,
where $\hat\threshold$ is the minimum cost of any codeword in the code
of cost at least $\threshold$.

\begin{algorithm}[h]
  \caption{--- {\em \proc{Round}: Construct a prefix-free code from a $\threshold$-relaxed code.}
  }
 \label{alg:RELAX} \label{alg:relax}
  \begin{algorithmic}[1]
    \smallskip
    \INPUT{$\threshold$-relaxed code $\code$ (for
      $\threshold=\lceil\log_2[\cost(1)/\eps] \cost(1)/\eps\rceil$).}
   \OUTPUT{Prefix-free code $\code'$ of cost $(1+O(\eps))\cost(\code)$}

    \STATE\label{step:binary}
    Define $\enc 0 =$ `00'.
    For integer $i>0$,
    define $\enc i$ to be the encoding of $i$
    obtained from the binary representation of $i$
    by replacing each `0' by `01', each `1' by `10',
    and finally appending `00'.
    \\Note that $\{\enc i : i=0,1,2,\ldots\}$ is prefix-free.
    \STATE Let $\hat\threshold = \min\{\cost(\code_k) : \cost(\code_k) \ge \threshold\}$.
    \FOR{each codeword $\code_k$ of cost $\threshold$ or more}
    \STATE
    {\bf Round} the codeword:
    let $x$ be the smallest prefix of $\code_k$ of cost $\threshold$ or more that is in $\universe$;
    let $y$ be the remaining suffix;
    replace the codeword $\code_k = xy$ by
    \,$\code'_k ~=~ x \enc {\cost(xy)-\hat\threshold}y$.
        \ENDFOR
    \STATE Return the rounded code $\code'$.
  \end{algorithmic}
  \end{algorithm}

\renewcommand{\reservedCode}{\code'}
\renewcommand{\relaxedCode}{\code}

Here is why the code $\reservedCode$ returned by \proc{Round} is prefix-free.
Since $\relaxedCode$ is $\threshold$-relaxed, codewords of cost less than $\threshold$
are not prefixes of any other codeword.
Any codeword of cost $i\ge \threshold$, once rounded,
cannot be a prefix of any non-rounded codeword
because the non-rounded codewords have cost less than $\threshold$.
It cannot be a prefix of any rounded codeword
because in any rounded codeword the string $\enc{i-\hat\threshold}$
(which immediately follows 
its unique minimal prefix $x$ of cost $\threshold$ or more in $\universe$)
uniquely determines the cost of the remaining suffix $y$.
Thus, $\reservedCode$ is prefix-free.

Here is why $\reservedCode$ has cost $(1+O(\eps))\cost(\code)$.
Modifying a codeword of cost $i\ge\threshold$ increases
its cost by at most $2\cost(1)\lceil\log_2 i\rceil$.
Since $i \ge \threshold$
and $\threshold$ is chosen\footnote{
The condition $\cost(1)\log \threshold = O(\eps \threshold)$
is equivalent to $\threshold/\log\threshold = \Omega(z)$ for $z=\cost(1)/\eps$.
This holds because the choice of $\threshold$ implies $\threshold \ge z\log z$,
which (using $\log z\le z$ and some algebra) implies $\threshold/\log\threshold \ge z/2$.
}
so that
$\cost(1)\log \threshold = O(\eps \threshold)$,
the increase is $O(\eps i)$.

Each modified codeword is still in $\universe$ because, 
in any codeword $xy$ that is modified,
the unmodified prefix $x$ is in $\universe$,
so $xz$ is in $\universe$ for any string $z$.
\end{proof}

\paragraph{Remark for intuition --- a slow PTAS}
Lemma~\ref{lemma:relax} alone is enough to give an $n^{O_\eps(1)}$-time PTAS
for \HUL (when $\cost(1)\le 3/\eps$).
The intuition is as follows.  

A minimum-cost $\threshold$-relaxed code $\code$ can be found as follows
(much more easily than a minimum-cost {\em prefix-free} code).
Let $\code_{< \threshold}$ denote the set containing the codewords in $\code$
of cost less than $\threshold$.
Given just $\code_{<\threshold}$,
the optimal way to choose the remaining codewords (those in $\code-\code_{< \threshold}$)
is {\em greedily}:
those remaining codewords must simply be 
some $n-|\code_{< \threshold}|$ {\em cheapest available strings
among those that have no prefix in $\code_{< \threshold}$}.
In short, the optimal $\threshold$-relaxed code $\code$ is essentially determined 
by its set $\code_{<\threshold}$ of codewords of cost less than $\threshold$.

In fact, the code $\code$ is essentially determined 
by just the {\em signature} $f$ of this set $\code_{<\threshold}$
(the signature $f$ essentially determines $\code_{<\threshold}$,
which in turn determines $\code$).
Each such signature is a distinct function $f:[\threshold]\rightarrow [0..n]$.
There are $(n+1)^\threshold$ such functions.

Recall that, as defined in Lemma~\ref{lemma:relax}, the threshold $\threshold$ is $O_\eps(1)$.
(The assumption $\cost(1) \le 3/\eps$
and the choice of $\threshold \approx \log[\cost(1)/\eps] \cost(1)/\eps$
imply $\threshold = O(\log(1/\eps)/\eps^2)$.)
Thus, the number $(n+1)^\threshold$ of such functions is $n^{O_\eps(1)}$.

The PTAS is as follows: exhaustively search all such functions $f$.
For each, construct a minimum-cost $\threshold$-relaxed code $\code$
such that $\code_{<\threshold}$ has signature $f$.
(If any such code $\code$ exists, 
it can be constructed greedily from just $f$ as described above.)
Finally, take $\code^{\min}$ to be the code of minimum cost
among the $\threshold$-relaxed codes $\code$ obtained in this way,
take $\code'$ to be the prefix-free code produced by $\proc{Round}(\code^{\min})$,
and, finally, return $\code'$.

\smallskip





By Lemma~\ref{lemma:relax}, 
the prefix-free code $\code'$ obtained by rounding $\code^{\min}$
has cost $(1+O(\eps))\cost(\code^{\min})$.
By its construction, $\code^{\min}$ is an optimal $\threshold$-relaxed code.
Since any prefix-free code is also $\threshold$-relaxed,
the cost of $\code^{\min}$ is at most
the cost of the minimum-cost prefix-free code, $\opt$.
Transitively,
$$\cost(\code') \le (1+O(\eps))\cost(\code^{\min}) \le (1+O(\eps))^2\opt = (1+O(\eps))\opt.$$
That is, the algorithm is a PTAS.

The rest of the paper is about reducing the running time
(in Sections~\ref{sec:cost1sig} and~\ref{sec:cost1tree})
and reducing the general case to the case $\cost(1)\le 3/\eps$
(in Section~\ref{sec:general}).

\subsection{Restricting to {\em group-respecting} $\tau$-relaxed codes}\label{subsec:grouping}
By Lemma~\ref{lemma:relax}, to find a near-optimal prefix-free code,
it suffices to find a near-optimal $\threshold$-relaxed code $\code$ and then ``round'' $\code$.

As described in the remark in Section~\ref{subsec:relax},
this fact yields a PTAS, one that works by exhaustively searching the potential
signatures $f$ for the set $\code_{<\threshold}$ of codewords of cost less than $\threshold$.
This gives an optimal $\threshold$-relaxed code $\code$,
which the PTAS then rounds to a near-optimal prefix-free code.

The run time of this PTAS is high because there are $n^{O_\eps(1)}$ potential signatures.

To reduce the run time, we next show how to compute a set $S$ of signatures
that has {\em constant} size yet is nonetheless still guaranteed to contain a good signature --- 
that is, the signature $f$ of some set $\code_{<\threshold}$
that extends to a near-optimal $\threshold$-relaxed code $\code$.

To compute this set $S$, we restrict attention to codes that choose the codewords
in levels less than $\threshold$ in a restricted way.
In particular, we partition the probabilities $\{\prob_i\}_i$ into a constant number of groups.
We then consider only codes that, within the levels less than $\threshold$,
{\em give all probabilities within each group codewords of equal cost}.

The partition $\group$ of $\prob[1..n]$ in question is constructed greedily
so that there are $O(\threshold/\eps) = O_\eps(1)$ groups, 
and, within each group,
either there is only one (large) probability
or the probabilities sum to $O(\eps/\threshold)$.
Recall that $\prob$ is decreasing.
\begin{definition}[grouping]\label{def:grouping}
Given any \HUL instance $(\prob,\alphabet,\cost)$, $\eps>0$,
and $\threshold$ from Lemma~\ref{lemma:relax},
define the {\em grouping}  $\group=\group_{\eps,\threshold}(\prob)$ of $\prob$
to be a partitioning of $\prob$'s index set $[n]$
into some $\gamma$ contiguous groups $(\group_1,\group_2,\ldots,\group_\gamma)$,
as follows:
take $\group_g = (j,{j+1},\ldots ,h)$,
where $h$ is maximal subject to~$\prob_j+\cdots+\prob_{h-1} \le \eps/\threshold$
(and $j$ is just after the previous group ended,
i.e. $j=1+\max \group_{g-1}$, or $j=1$ if $g=1$).

\smallskip
Given a $\threshold$-relaxed code $\code$, say $\code$ {\em respects $\group$} if,
for each group $\group_g$, if any index $k$ in $\group_g$ is assigned a codeword 
of some cost $i$ less than $\threshold$, then {\em all} indices in $\group_g$ are assigned codewords of cost $i$.
(Formally, for all $g$, for any $k,k'\in\group_g$, one has
$\max(\cost(\code_k), \threshold) = \max(\cost(\code_{k'}), \threshold)$.)
\end{definition}\smallskip

The number of groups, $\gamma$, is at most $\threshold/\eps$
(because each group except the last has total probability at least $\eps/\threshold$).
Also, each group $G_g=(j,j+1,\ldots,h)$ either has just one member, 
or has $\prob_j + \prob_{j+1} + \prob_{h-1} \le \eps/\threshold$.

Next we argue that there is always a $\group$-respecting $\threshold$-relaxed 
code that is a near-optimal.
To do this, we show that any $\threshold$-relaxed code (in particular the optimal one)
can be modified, by working from level 0 to level $\threshold-1$,
appending `0's to codewords as necessary
to make the code $\group$-respecting, while increasing the cost 
by at most a $1+\eps$ factor.
More specifically, since the code is monotone, in any given level $i<\threshold$,
at most one group $\group_g$ is ``split'' between that level and higher levels,
and that group has total probability $O(\eps/\threshold)$.
We ``fix'' that group (by appending a `0' to its level-$i$ codewords)
while increasing the cost of the code by $O(\cost(0)\eps/\threshold)$.
The total cost of fixing all levels in $[0,\threshold-1]$ in this way
is at most $\threshold\times\cost(0)\eps/\threshold = \cost(0)\eps$.
This is at most $\eps$ times the total cost of the code,
because any code must cost at least $\cost(0)$.

\begin{lemma}[grouping gap]\label{lemma:group}
Given a $\threshold$-relaxed code $\code$ for any \HUL instance,
there exists a 
$\threshold$-relaxed code $\code'$ that is $\group$-respecting and 
such that $\cost(\code')\le (1+\eps)\cost(\code)$.
\end{lemma}

\begin{proof}
Let $\code$ be any $\threshold$-relaxed code.
If $\code$ is not monotone, reorder its codewords to make it monotone.
\begin{figure*}[t]
\centering
\includegraphics[width=\textwidth]{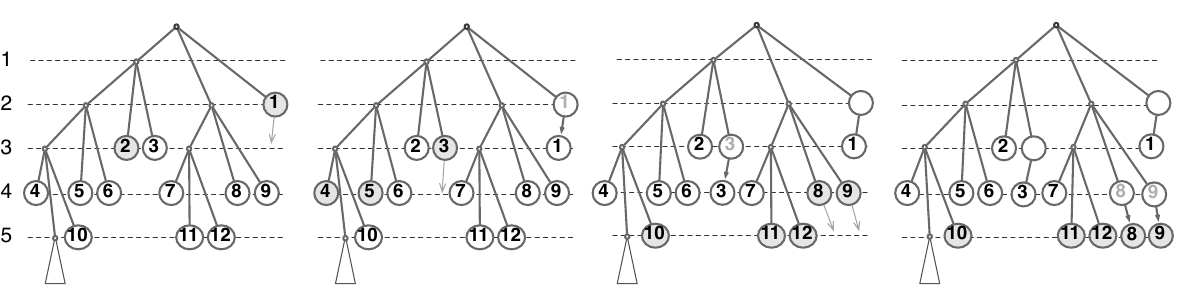}
\caption{\em Making a code $\group$-respecting.
The first four groups are $\group_1=\{1,2\}, \group_2=\{3,4,5\},
\group_3=\{6,7\}, \group_4=\{8,9,10,11,12\}$.
Iterations for levels $i=2,3,4$ are shown, from left.}
\label{fig:group}
\hrulefill
\end{figure*}
For each $i\in[\threshold]$, in increasing order, do the following.
Since $\code$ is monotone there can be at most one group $\group_g$ that is ``split''
at level $i$, meaning that some probabilities are assigned codewords of cost $i$ 
while others are assigned codewords of larger cost.
If there is such a group, add a letter `0'
to the end of each level-$i$ codeword assigned to that group,
and then reorder the codewords above level $i$  to restore monotonicity.
This defines $\code'$.

Note that the codewords in $\code'$ are still in $\universe$, and that
$\code'$ is monotone, $\group$-respecting, and $\threshold$-relaxed.

To finish we bound the cost increase.
Clearly, reordering codewords to make a code monotone never increases the cost.
Then, if a group $\group_g=(j,j+1,\ldots,h)$ has its codewords modified for level $i$, 
then that group must have at least two members,
and $\prob_j + \prob_{j+1} + \cdots + \prob_{h-1}$ must be at most $\eps/\threshold$.
Thus, adding a letter `0' to the level-$i$ codewords assigned to $\group_g$ 
increases the cost of the code by at most $\cost(0)\eps/\threshold$.
Since there is at most one such increase  for each level $i<\threshold$,
the total increase in cost is at most $\threshold\cost(0) \eps/\threshold = \eps\cost(0)$.
On the other hand, the cost of any code is at least $\cost(0)$.
Thus, the modified code $\code'$ has cost at most $(1+\eps)\cost(\code)$.
\end{proof}

\subsection{Bounding the support of $\tau$-relaxed group-respecting codes}\label{subsec:limitedlevels}
By Lemma~\ref{lemma:group}, to find a near-optimal $\threshold$-relaxed code,
it suffices to find a near-optimal $\group$-respecting $\threshold$-relaxed code $\code$.

In this section, we observe that any such code $\code$
(and its prefix-free rounded code $\code'$ per Lemma~\ref{lemma:relax}) must have {\em support}
in a logarithmic-size set $\limitedlevels$ of levels.
That is, each string in $\prefixes(\code)\cap\universe$ 
(and each node in its tree representation)
must have cost in $\limitedlevels$.
Thus, for example, the signature $x$ of such a code has support of logarithmic size.
\smallskip

We use this structural property later in the paper to keep parts of the computation time poly-logarithmic.
The detailed definition of $\limitedlevels$ is not important;
what is important is that $\limitedlevels$ can be precomputed easily 
and has logarithmic size.

\begin{definition}[limited levels, $\limitedlevels$]\label{def:limitedlevels}
Given any instance of \HULR,
let $\threshold$ be as defined in Lemma~\ref{lemma:relax}.
Let $i_\roots$ be the minimum cost of any root of $\universe$ of cost at least $\threshold$.
Let  $i_\alphabet$
be the minimum cost of any letter in $\alphabet$ of cost at least $\threshold$.
Let $\delta = \cost(1)\lceil \log_2 n\rceil$.
Define $\limitedlevels$, the {\em set of possible levels}, to contain the 
$O(\poly(\eps^{-1})\log n)$ integers in
\begin{equation}
  \label{eq:limitedlevels}
  [0,2\threshold+3\delta]
  ~\cup~ [i_\roots, i_\roots + 3\delta] 
  ~\cup~ [i_\alphabet, i_\alphabet + \threshold + 3\delta].
\end{equation}
 (If $i_\roots$ or $i_\alphabet$ is not well defined, take the corresponding interval above to be empty.)
\end{definition}\smallskip

To verify that $\limitedlevels$ has logarithmic size,
note that, since $\cost(1) \le 3/\eps$, it follows that $\threshold = O(\poly(\eps^{-1}))$
and $\delta = O(\poly(\eps^{-1}) \log n)$.
Thus, by inspection, $\limitedlevels$ has size $O(\poly(\eps^{-1})\log n)$.

Next we prove that without loss of generality,
in computing and rounding a $\threshold$-relaxed code,
we can limit attention to codes having support in $\limitedlevels$.

\begin{wrapfigure}{r}{0.4\textwidth}
   \includegraphics[width=0.4\textwidth]{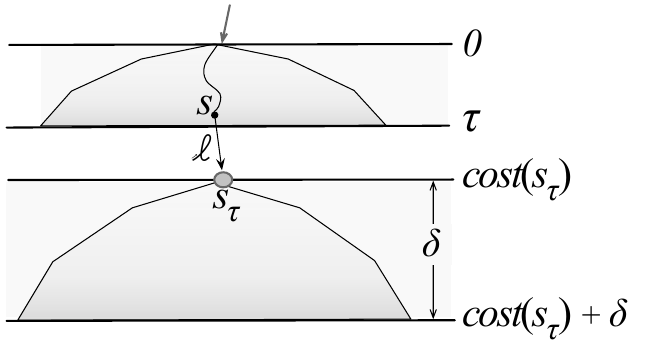}
\end{wrapfigure}
The proof is based on local-optimality arguments
(and details of the rounding procedure).
The rough idea
is this. 
Among the words in levels $\threshold$ and up that are
available to be codewords, 
let $s_{\threshold}$ denote one of minimum cost, as shown to the right.
Since codewords in levels $\threshold$ and above
must be taken greedily in any optimal
$\threshold$-relaxed code, 
and the $n$ words of the form $s_{\threshold}\{0,1\}^{\lceil \log_2 n\rceil}$
are available to be codewords, it follows that all codewords
that lie in level $\threshold$ or above should have costs
in $[\cost(s_{\threshold}), \cost(s_{\threshold})+\delta]$ (recall $\delta=\cost(1)\lceil \log_2 n\rceil$).
To finish the proof, we bound the values that $\cost(s_{\threshold})$ can take,
and we observe that rounding any codeword in level $\threshold$ or above
increases its costs by at most $2\delta$.

\begin{lemma}[limited levels]\label{lemma:limitedlevels}
Given any instance of \HULR,
let $\limitedlevels$ be as defined above.

(i) Any minimum-cost $\threshold$-relaxed, $\group$-respecting code $\code$
has support in $\limitedlevels$.

(ii) Rounding such a code $\code$ (per Lemma~\ref{lemma:relax})
gives a prefix-free code $\code'$ with support in $\limitedlevels$.
\end{lemma}

\begin{proof}
\noindent {\em Part (i).}
Let $\code$ be any minimum-cost $\group$-respecting $\threshold$-relaxed code.
Assume $\code$ has a codeword of cost at least $\threshold$
(otherwise all nodes in the tree representation 
are in $[0,\threshold) \subset \limitedlevels$, and we are done).

Say a string of cost at least $\threshold$ is {\em available} 
if no prefix of the string is a codeword of cost less than $\threshold$ in $\code$.

Let $s_\threshold$ be a minimum-cost available codeword.
(There is at least one, by the assumption that $\code$ has a codeword of cost at least $\threshold$.)
Let $s$ be the parent of $s_{\threshold}$, so that $s_{\threshold} =
s\ell$ for some $\ell\in\alphabet$, as shown in the figure above. 

The $n$ strings in $S = s_\threshold\{0,1\}^{\lceil \log_2 n\rceil}$ are available.  
Each costs at most $\cost(s_{\threshold})+\delta$,
so, in the tree representation of $\code$,
all levels $i> \cost(s_\threshold)+\delta$ are empty 
(otherwise $\code$ could be made cheaper by swapping in
some string of cost at most $\cost(s_{\threshold})+\delta$).
Thus, $\code$ has support in
\(
[0,\threshold) \cup [\cost(s_\threshold), \cost(s_\threshold) + \delta].
\)

Let $\code'$ be obtained by rounding $\code$ (Lemma~\ref{lemma:relax}).
Any unmodified codeword has cost less than $\threshold$.
Following the notation of Alg.~\ref{alg:relax},
let $\code'_k = x\enc{i-\cost(s_\threshold)}y$ be any modified codeword,
so that $i=\cost(\code_k)$.
By the previous paragraph, $i\le \cost(s_\threshold)+\delta$,
and rounding increases the cost of the codeword
by at most $\cost(\enc{\delta}) \le 2\cost(1)\lceil \log_2 \delta \rceil \le 2\delta$
(assuming $n\ge 3/\eps$)
to at most $\cost(s_{\threshold})+3\delta$.
Also, by the rounding method, the code tree is not modified below level $\cost(s_\threshold)$.
Thus, $\code'$ and $\code$ have support in
\(
[0,\threshold) \cup [\cost(s_\threshold), \cost(s_\threshold) + 3\delta].
\)

To complete the proof, we show that these two intervals are contained within the
three intervals 
$[0,2\threshold+3\delta]
\cup [i_\roots, i_\roots + 3\delta] 
\cup [i_\alphabet, i_\alphabet + \threshold + 3\delta]$
from the definition of $\limitedlevels$.
By inspection, this will be the case as long as 
\begin{equation}
  \label{eq:sthresh}
  \cost(s_{\threshold}) ~\in~
  [\threshold,2\threshold] \cup \{i_\roots\} \cup [i_\alphabet, i_\alphabet+\threshold].
\end{equation}
We use a case analysis to show that (\ref{eq:sthresh}) holds.

If it happens that $s\not\in\universe$, then $s_{\threshold}$ is a root of $\universe$, 
necessarily (by the choice of $s_{\threshold}$) of cost $i_\roots$, 
so (\ref{eq:sthresh}) holds.
So assume $s\in\universe$.
Then $\cost(s)<\threshold$ (otherwise $s$ would be available and have
cost less than $s_{\threshold}$, contradicting the choice of $s_\threshold$).
If it happens that $\cost(\ell)<\threshold$,
then $\cost(s_{\threshold}) = \cost(s)+\cost(\ell) < 2\threshold$,
so $\cost(s_\threshold) \in [\threshold,2\threshold]$, and (\ref{eq:sthresh}) holds.
So assume $\cost(\ell)\ge \threshold$.
In this case $i_\alphabet$ is well-defined and $\cost(\ell)\ge i_\alphabet$
(as no letters have cost in $[\threshold, i_\alphabet)$, by the definition of $i_{\alphabet}$).
In fact it must be that $\cost(\ell) = i_\alphabet$
(otherwise replacing the last letter $\ell$ in codeword $s_{\threshold}$
by the letter of cost $i_\alphabet$ would give a string that is cheaper than $s_\threshold$,
contradicting the choice of $s_\threshold$).
Thus, $i_\alphabet \le \cost(\ell) \le \cost(s_\threshold) \le \threshold+i_\alphabet$,
so $\cost(s_\threshold)\in[i_\alphabet, i_\alphabet+\threshold]$.
\end{proof}





%





%

\subsection{A mixed integer program to find a min-cost $\group$-respecting $\threshold$-relaxed code}\label{subsec:MP}

In this section we focus on the problem of finding
the full signature $x$
of an optimal
$\group$-respecting 
$\threshold$-relaxed
code
$\code$,
for a given instance of \HUL.
We describe how this problem can by modeled 
by an integer linear program ($\MPG$)
that (thanks to Lemma~\ref{lemma:limitedlevels})
has size $O_\eps(\log n)$,

We also identify, within $\MPG$,
a particular constant-size vector $z$ of binary variables.
(These variables encode the assignment of the groups in $\group$
to the levels less than $\threshold$.)
We show that, given any assignment to just these constantly many binary variables,
an optimal assignment of the remaining variables 
can be computed greedily in $O_\eps(\log^2 n)$ time.
Thus, by exhaustive search over the $O_\eps(1)$ possible assignments to $z$,
one can find an optimal solution to $\MPG$ 
(and hence the signature $x$ of an optimal $\group$-respecting $\threshold$-relaxed code)
in $O_\eps(\log^2 n)$ time.

The integer linear program
$\MPG$ is a modification of one of Karp's original integer programs \cite[\S IV]{Karp-61} for \HUL
(that is, for finding a minimum-cost {\em prefix-free} code;
in contrast we seek a $\group$-respecting, $\threshold$-relaxed code).
The variables of $\MPG$ are contained in four vectors $(w,x,y,z)$,
where $x$ encodes the signature of the codeword set,
$w$ encodes the signature of the set of interior nodes,
$y$ encodes the assignment of probabilities to levels
($y$ is determined by $x$, and helps compute the cost),
and $z$ encodes the assignment of groups to levels 
(for levels less than $\threshold$).
The basic idea (following Karp) 
is that, since the numbers of various types of nodes available on level $i$
satisfy natural linear recurrences in terms of the numbers at lesser levels,
we can model the possible signatures 
by linear constraints on $x$ and $w$.

For intuition, 
we first describe Karp's original integer program for finding a {\em prefix-free} code
(generalized trivially here to allow a universe $\universe$ with arbitrary root set $\roots$).
The inputs to Karp's program
are the probability distribution $\prob$ along with the signatures $\sigma$ and $r$
of, respectively, the alphabet $\alphabet$ and the root set $\roots$.
(Note that $m = n\max\{\cost(\ell)~|~\ell\in\alphabet\}$
is a trivial upper bound on any codeword cost in any optimal code.)
Karp's program is in Fig.~\ref{fig:karp}.
\begin{figure}[t]
\begin{LPTAB}
\begin{LP}{\MP}
 x_i + w_i & \le & r_i+\displaystyle\sum_{j<i} \sigma_{i-j} w_j & (i \in [m]) 
\\
 \sum_{k\in[n]} y_{ki} & = & x_i & (i \in [m])
\\
 \sum_{i\in[m]} y_{ki} & = & 1 & (k\in[n])
\smallskip
\\
 w_i,x_i,y_{ki} & \in  & \Nz & (i \in [m],
\\
&&& ~~k\in[n])
\end{LP}
&
~\parbox[t]{0.47\textwidth}{\small
\begin{tabular}[t]{@{~}c@{}@{ --- }l}
\multicolumn{2}{c}{\em --- ~ Parameters of $\MP$ ~ ---}\\[2pt]
$\prob$ & probability distribution on $[n]$ \\
$r$ & signature of $\universe$'s root set $\roots$\\
$\sigma$ & signature of alphabet $\alphabet$\\
$m$ & $n\times\max\{\cost(\ell)~|~\ell\in\alphabet\}$ 
\\[4pt]
\multicolumn{2}{c}{\em --- ~ Variables determining code $\code$ ~ ---} \\[2pt]
$x$ & signature of codewords ($\code$)  \\
$w$ & signature of interior nodes \\
$y$ & assignment; $y_{ki}=1$ iff $\cost(\code_k) = i$ \smallskip
\end{tabular}
}
\end{LPTAB}
\caption{Karp's integer linear program for finding a minimum-cost prefix-free code.}\label{fig:karp}
\end{figure}

We call the first constraint in $\MP$ the ``capacity'' constraint.
Note that the vector $z$ is not used in $\MP$.

\begin{theorem}[Correctness of $\MP$, \cite{Karp-61}, \S IV]\label{thm:karp61}
In any optimal solution $(w^\smallopt, x^\smallopt, y^\smallopt)$ of $\MP$,
the vector $x^\smallopt$ is the signature of a minimum-cost prefix-free code,
the cost of which is the cost of $(w^\smallopt, x^\smallopt, y^\smallopt)$.
\end{theorem}\smallskip

{\em Proof sketch.}
For any prefix-free code $\code$, there is a feasible solution $(w,x,y)$ 
for $\MP$ of cost $\cost(\code)$.
To see why, consider the tree representation of $\code$.
Let $x_i$ be the number of leaves in level $i$,
let $w_i$ be the number of interior nodes (in $\universe$) in level $i$,
and let $y_{ki}=1$ if $\cost(\code_k) = i$, and $y_{ki}=0$ otherwise.
(So $y_{ki}$ indicates whether probability $\prob_k$ is assigned to level $i$.)
Taking $(w,x,y)$ as a solution to $\MP$,
the capacity constraint holds because each interior node
on level $j$ can have at most $\sigma_{i-j}$ children in level $i$.
By inspection, the other constraints are also met,
and $(w,x,y)$ has cost equal to $\cost(\code)$.

Conversely, given any feasible solution $(w,x,y)$,
one can greedily construct a code $\code$
with signature $x$ by building its tree representation level by level
(in order of increasing $i\in\limitedlevels$),
adding $w_i$ interior nodes and $x_i$ codeword nodes in level $i$.
The capacity constraint ensures that there 
are enough parents (and roots) to allocate each level's nodes.
\endproof

Next we modify $\MP$ to model our problem:
finding the signature $x$ of a minimum-cost 
{\em $\group$-respecting
$\threshold$-relaxed} code
(instead of a minimum-cost prefix-free code).
The modified program, denoted $\MPG$,
is shown in Fig.~\ref{fig:MPG}
The program differs from Karp's
in three ways, labeled (a), (b), (c).

\begin{figure}[t]
\begin{LPTAB}
\begin{LP}{\MPG}
\begin{array}{rr}
\text{\em if } i<\threshold: & x_i + w_i 
\\
\hspace*{-3.1em}{\text{\small\em \parbox{1em}{(a)}}\rightarrow} \hfill
\text{\em if } i\ge\threshold: & \max(x_i,w_i)
\end{array}
\Big\}
 & \le & r_i + \displaystyle\sum_{j<i} \sigma_{i-j}w_j & (i \in \limitedlevels) 
\hfill ~~\lefteqn{\leftarrow (b)}
\\
 \sum_{k} y_{ki} & = & x_i & (i \in \limitedlevels)
\\
 \sum_{i} y_{ki} & = & 1 & (k\in[n])
\smallskip
\\
\hspace*{-2.6em}{\text{\small\em \parbox{1em}{(c)}}\rightarrow} \hfill
 z_{gi} & \in  & \{0,1\} & (i \in [\threshold-1], g\in[\gamma])
\\
y_{ki} & =  & z_{gi} & (i \in [\threshold-1], g\in[\gamma],k\in \group_g)
\smallskip
\\
 w_i,x_i,y_{ki} & \in  & \Nz & (i \in \limitedlevels, k\in[n])
\end{LP}
\end{LPTAB}
\caption{An integer program for computing an optimal $\threshold$-relaxed,
  $\group$-respecting code.}\label{fig:MPG}
\end{figure}

\begin{itemize}
\item[\em (a)] {\em For $i$ above the threshold $\threshold$, the left-hand side of the capacity constraint
is replaced by $\max(x_i,w_i)$.}

This models $\threshold$-relaxed codes,
in which codeword nodes in level $i\ge \threshold$ can also be interior nodes.

\item[\em (b)] {\em The indices $i$ (and $j$) range over the set $\limitedlevels$ of possible levels,
instead of $[m]$ (per Defn.~\ref{def:limitedlevels}).}

Restricting $i$ and $j$ to levels within $\limitedlevels$
is without loss of generality by Lemma~\ref{lemma:limitedlevels}.

\item[\em (c)] {\em There are $\threshold\gamma$ new 0/1 variables: one variable $z_{gi}$ 
for each group $\group_g$ ($g\le \gamma$) and  level $i<\threshold$.}


\end{itemize}

The new $z$ variables enforce the restriction to $\group$-respecting codes.
Specifically, they constrain the $y$ variables to force all probabilities 
within a given group to be assigned to the same level
(if any is assigned to a level below $\threshold$):
$z_{gi}$ will be 1 iff group $\group_g$ is assigned to level $i<\threshold$
(if a group is not assigned to any level below $\threshold$, then all its 
$z_{gi}$'s will be zero).
\smallskip

Next we state the formal correctness of $\MPG$:
that the feasible solutions to $\MPG$
do correspond to the (signatures of the) $\group$-respecting $\threshold$-relaxed codes.

\begin{lemma}[correctness of $\MPG$]\label{lemma:hulg code to solution}
(i) Given any minimum-cost $\threshold$-relaxed $\group$-respecting code $\code$,
the integer program $\MPG$ has a feasible solution $(w,x,y,z)$
of cost $\cost(\code)$ where $x$ is the signature of $\code$.

\label{lemma:hulg solution to code}
(ii) Conversely, given any solution $(w,x,y,z)$ of $\MPG$,
there is a $\threshold$-relaxed $\group$-respecting code $\code$ 
having signature $x$ and with equal (or lesser) cost.
\end{lemma}

{\em Proof sketch.}
(A detailed proof is in the Appendix.)

The proof is a simple extension of  the proof of Theorem~\ref{thm:karp61}.
In the forward direction,
the capacity constraint is met because, in any $\threshold$-relaxed code,
codeword nodes in levels $\threshold$ and higher can also be interior nodes.
In the backward direction,
the code 
is $\group$-respecting because
of the constraint
$y_{ki} = z_{gi}$
(for 
$g\le\gamma$,
$k\in \group_g$,
and $i<\threshold$).
\endproof

\paragraph{Remark}
We remark without proof that the integrality constraints on $w$, $x$, and $y$
(in the final line of $\MPG$) can be dropped, giving a mixed integer linear program.
(In any optimal basic feasible solution to the latter program, $w$, $x$, and $y$ will still take only
integer values.)
\smallskip

Note that a particular assignment of the $z$ variables 
determines the assignment of groups in $\group$ 
within each level in $[0,\threshold-1]$.
As previously discussed,
this in turn essentially determines the rest of the $\threshold$-relaxed code,
as codewords in levels $\threshold$ and above should be chosen greedily.
Thus, given any particular assignment of the variables in $z$,
there is a natural optimal assignment of the remaining variables $(w,x,y)$.
We call this $(w,x,y,z)$ the {\em greedy extension} of $z$.
Here is the formal definition.

\begin{definition}[greedy extension]\label{def:hulg extension}
Given any $z$ with values in $\{ 0,1\}$ such that $\sum_i z_{gi} \le 1$ for each $g$,
define the {\em greedy extension} of $z$ for $\MPG$ 
to be the tuple $(\hat w,\hat x,\hat y,z)$ of all-integer vectors
defined as follows:

1. In each level $i<\threshold$, in increasing order, 
  define $\hat x_i$ and $\hat w_i$ as follows.
  Let $\hat x_i$ be the number of probabilities that $z$ assigns to level $i$,
  that is, $\hat x_i = \sum_{g: z_{gi}=1} |\group_g|$. 
  Let $\hat w_i$ be the number of interior nodes left available in level $i$.
  That is, let $\hat w_i$ be maximal subject to the capacity constraint.

2. For each level $i\ge \threshold$, in increasing order, 
take interior and codeword nodes greedily:
take $\hat x_i$ and $\hat w_i$ to be maximal
subject to the capacity constraint for $i$ and the constraint $\sum_{j\le i} \hat x_j \le n$.

3. Among vectors $y$ such that the tuple $(\hat w, \hat x, y, z)$ is feasible for $\MPG$,
  let $\hat y$ be one giving minimum cost (breaking any ties by assigning 
  probabilities with lesser indices to lesser levels).

\end{definition}\smallskip
Note: In Step 1, if it happens that
the capacity constraint is violated even with $\hat w_i  = 0$,
then there is no $\group$-respecting $\threshold$-relaxed code
for the given $z$, and the greedy extension of $z$ is not well-defined.

In Step 2, if it happens that
some probabilities are not assigned to any level below $\threshold$
(i.e.,  $\sum_{i<\threshold} \hat x_i < n$)
but no nodes are available in higher levels 
(i.e., for all $i\ge \threshold$, the right-hand side of the capacity constraint is 0),
then there is no $\group$-respecting $\threshold$-relaxed code
for the given $z$, and the greedy extension of $z$ is not well-defined.
\smallskip


Since codewords in levels $\threshold$ and higher should be assigned greedily,
the greedy extension is optimal:
\begin{lemma}[optimality of greedy extension]\label{lemma:hulg extension}
Fix any $z$ for which there is any feasible extension $(x,w,y,z)$ for $\MPG$. 
Then the greedy extension $(\hat w,\hat x,\hat y,z)$ of $z$ is well-defined, 
feasible, and has minimum cost.
\end{lemma}

The proof is straightforward; it is in the appendix.

The next corollary summarizes what is needed from this section:
\begin{corollary}[correctness of $\MPG$]\label{cor:mpg}
Fix any instance of \HUL.

\smallskip
\noindent{\em (i)} Fix any $z$ that has some feasible extension for $\MPG$. 
Then the greedy extension $(\hat w,\hat x,\hat y,z)$ of $z$ 
is well-defined, feasible, and has minimum cost.

\smallskip
\noindent{\em (ii)} Let $(w^\smallopt, x^\smallopt, y^\smallopt, z^\smallopt)$ be an optimal solution to $\MPG$.
Then $x^\smallopt$ is the signature of a minimum-cost $\group$-respecting $\threshold$-relaxed code.
\end{corollary}

Part (i) of the corollary is just Lemma~\ref{lemma:hulg extension}.
Part (ii) follows from Lemma~\ref{lemma:hulg code to solution}.


\subsection{Proof of Theorem~\ref{thm:cost1sig}}
\label{subsec:cost1sig proof}
We now prove Thm.~\ref{thm:cost1sig}:

\thmcostsig
\begin{proof}
  By Lemmas~\ref{lemma:relax}--\ref{lemma:limitedlevels} and Corollary~\ref{cor:mpg},
  the steps in Fig.~\ref{fig:PTAS} give the signature $x'$ and cost.
 \begin{figure}[t]
    \centering
 \parbox{0.95\textwidth}{\em \setlength{\parskip}{1ex}
    \hrulefill

  0. Let $\threshold=\lceil\cost(1)\log_2(\cost(1)/\eps)/\eps\rceil$.

1. \label{step:1} Compute grouping $\group=\group(\prob)$ (Defn.~\ref{def:grouping})
and set of levels $\limitedlevels$ (Defn.~\ref{def:limitedlevels}).

2. \label{step:2} For each possibly feasible assignment $\hat z$ to $z$ in $\MPG$:

~2a. Compute just $\hat w$ and $\hat x$ 
	of the greedy extension of $\hat z$ (Defn.~\ref{def:hulg extension}). 

~2b. From $\hat x$, compute the cost of the greedy extension of $\hat z$
          (if well defined).

Select $(w^\smallopt,x^\smallopt,z^\smallopt)$ to be the $(\hat w, \hat x, \hat z)$ 
giving min.~cost among those computed.

3. \label{step:3} Without explicitly computing the
  $\threshold$-relaxed code $\code$ with signature $x^\smallopt$, compute\\ the
  signature $x'$ and approximate cost of the prefix-free code $\code' =
  \proc{Round}(\code)$.

\hrulefill
}
    \caption{The steps of the PTAS for the case $\cost(1)\le 3/\eps$.}\label{fig:PTAS}
\end{figure}

To finish, we show that each of these steps can be
done in $O_\eps(\log^2 n)$ time, given $\PROB$, $\sigma$, and $r$.

\paragraph{Step~1}
Compute $G$ (in particular, the first and last index of each group $\group_g$) as follows.
By inspection of Defn.~\ref{def:grouping},
for each group $\group_g=(j,..,h)$,
the index $h$ can be computed 
in $O(\log n)$ time from $\PROB$ by binary search.
There are  at most $\threshold/\eps$ groups,
so the total time is $O( (\threshold/\eps) \log n) = O_\eps(\log n)$.

Compute $\limitedlevels$ in time $O(|\limitedlevels|) = O_\eps(\log n)$ as follows.
Following Defn.~\ref{def:limitedlevels}, compute $i_\roots$
and  $i_\alphabet$ in $O(\threshold)=O_\eps(1)$ time
(assuming $r$ and $\sigma$ are given as sorted lists or arrays indexed by $i$),
then enumerate $\limitedlevels$.

\paragraph{Step~2}
There are at most $\threshold^{|\group|} = O_\eps(1)$ possibly feasible assignments to $z$.
(An assignment chooses a level in $[\threshold-1]$, or no such level, for each group index $g\in[\gamma]$;
although $\MPG$ allows other assignments to $z$ in which $\sum_{i} z_{gi} > 1$,
none of those will have a feasible extension because they force $\sum_i y_{ki} > 1$ for $k\in\group_g$.)

For each such assignment $\hat z$, to compute just $\hat w$ and $\hat x$ 
of the greedy extension (Defn.~\ref{def:hulg extension}), observe that
all $\hat x_i$ with $i<\threshold$ can be set in total time $O(|\group|) = O(\gamma) = O_\eps(1)$ using $\hat x_i = \sum_{g : \hat z_{gi}=1} |\group_g|$.
Then, the $\hat w_i$ (for $i\in\limitedlevels$), 
and the $\hat x_i$ (for $i\in \limitedlevels, i\ge \threshold$), 
can each be computed in time $O(|\limitedlevels|)$ 
(the time it takes to compute $\sum_{j<i}\sigma_{i-j} w_j$),
for a total time of $O(|\limitedlevels|^2) = O_\eps(\log^2 n)$.

Given $\hat z$ and $\hat x$, the cost of the code 
can then be computed (without computing $y$!) as follows.
The probability associated with a group $\group_g$
is $\PROB[\max\group_g] - \PROB[\max\group_{g - 1}]$.
The contribution of levels less than $\threshold$ to the cost
is $\sum_{g}\sum_{i<\threshold} i\, \hat z_{gi} (\PROB[\max\group_g] - \PROB[\min\group_g - 1])$.

The cumulative cost of codewords in levels $i\ge\threshold$ can be computed as follows.
Consider those groups $\group_g$ that are not assigned to the lower levels,
in order of increasing $g$.
Break the groups as necessary into smaller pieces,
while assigning the pieces monotonically to the levels $i=\threshold, \threshold+1, \ldots$,
so that each level $i$ is assigned pieces of total size $x_i$.
(At most $|\group|+|\limitedlevels|-\threshold$ pieces will be needed to do this.)
Once all pieces are assigned levels,
compute their cumulative cost as the sum, over the pieces,
of the cumulative probability in the piece times the assigned level.
In this way, the cost of the code for a given $\hat z$ and $\hat x$
can be computed in time $O(|G|\threshold + |G|+|\limitedlevels|) = O_\eps(\log n)$.

Since there are $O_\eps(1)$ assignments $\hat z$ to consider,
and for each $\hat x$ can be computed in $O_\eps(\log^2n)$ time,
the total time to find the minimum-cost signature $x$ is $O_\eps(\log^2 n)$.

\paragraph{Step~3}
By inspection of \proc{Round} in the proof of Lemma~\ref{lemma:relax},
for each codeword of cost $i\ge\threshold$ in $\code$,
there is a codeword of cost $i+\cost(\enc{i-\hat\threshold})$ in $\code'$.
Thus, $x'$ can be computed directly from $x$ by taking $x'_i = x_i$ for $i<\threshold$,
and for the rest, starting with $x'_i = 0$ and then, for each $i\ge\threshold$,
incrementing $x'_{i'}$ by $x_i$ where $i' = i + \cost(\enc{i-\hat\threshold})$.

The cost of $\code'$ is $1+O(\eps)$ times the cost
of the $\threshold$-relaxed code with signature $x$,
which is, in turn, the cost of the solution $(w,x,y,z)$ to $\MPG$,
which is known from the previous step.

This completes the proof of Thm.~\ref{thm:cost1sig}.
\end{proof}
\smallskip

The following observations about the proof are useful in the next section.
By 
Lemma~\ref{lemma:limitedlevels},
the code whose signature is produced has support in $\limitedlevels$.
Thus, the tree representation uses only the roots of $\universe$ that lie in levels in $\limitedlevels$.  Similarly, by inspection of $\MPG$, its solution
requires only those $r_i$ with $i\in\limitedlevels$.  In sum:

\begin{subobservation}\label{obs:limitedlevels}
The computation in Thm.~\ref{thm:cost1sig}
produces a signature $x$ for a code with support in $\limitedlevels$.
The computation does not require the full signature $r$ of the roots of $\universe$,
but relies only on the $r_i$ such that $i\in\limitedlevels$
(the set $\limitedlevels$ of possible levels from Defn.~\ref{def:limitedlevels}).
\end{subobservation}

\section{Computing the tree representation from the signature}
\label{sec:cost1tree}
\newcommand{\MPE}{\text{\sc Edges}\ensuremath{_\limitedlevels(x,r,\sigma)}}

For the case $\cost(1)\le 3/\eps$,
Thm.~\ref{thm:cost1sig} proves that the signature (and cost) of a near-optimal
prefix-free code can be efficiently computed, but says nothing about computing a more explicit
representation of the code.  Here we address this by proving Thm.~\ref{thm:cost1tree}, 
which describes how to compute the tree representation 
in $O(n)+O_\eps(\log^2 n)$ time, given the signature $x$.



Given the signature, it would be easy to compute the tree-representation $F$ using a root-to-leaves
greedy algorithm in time $O(|F| + |\limitedlevels|)$ 
(where $|F|$ is the number of nodes in $F$).
Roughly, one could just allocate the nodes and edges of $F$ appropriately
in order of increasing level $i\in\limitedlevels$.
Unfortunately, $F$ might not have size $O(n)$, because in the worst case 
it may have many long chains of interior nodes each with just one child.\footnote
{Indeed, for some instances, there are signatures that force this to happen.}

One could of course modify $F$, splicing out nodes with just one child, so as to build a new tree $F'$ whose size is $O(n)$ and whose cost is less than or equal to the cost of $F$. 
However, if the algorithm were to explicitly build $F$ from the signature, and then modify $F$ into $F'$ as described, it would still take time at least $O(|F|)$, which could be excessive.
To prove the theorem below, we describe how to bypass the intermediate construction of $F$,
instead building $F'$ directly from $x$, in time $O(|F'|)+ O_\eps(\log^2 n)$, where $|F'|=O(n)$.


\begin{theorem}\label{thm:cost1tree}
Given any instance $\instance=(\prob,\alphabet,\cost,\universe)$ of \HUL with restricted universe such that $\cost(1) \le 3/\eps$,
and given the signature $x$ of some prefix-free code $\code$ with support in $\limitedlevels$,
one can construct the tree representation of a prefix-free code $\code'$
that has cost at most $\cost(\code)$. The running time is $O(n) + O_\eps(\log^2 n)$.
The tree representation has $O(n)$ nodes.
\end{theorem}\smallskip

\begin{proof}
Starting from the signature $x$,
we first compute various signatures for a tree $F$ whose codeword nodes have signature $x$.
Specifically, we compute both $w$ (the signature of the interior nodes of $F$)
and an ``edge signature'' $e$ --- where $e_{ji}$ is the number of edges from level $j$
to level $i>j$ in $F$. 
In fact the signature $x$ does not uniquely determine $e$ or $w$, so we make some arbitrary choices
to fix a particular $F$ with codeword signature $x$.

Here are the details of how to compute $w$ and $e$ in time $O(|\limitedlevels|^2)$. 
%

1. To start, initialize vector $w$ so that
the capacity constraint for $\MP$ (on the left below) holds with $x$:
\smallskip
\begin{LPTAB}
\begin{LPNoCost}{the capacity constraint for $\MP$}
 x_i + w_i & \le & r_i+\displaystyle\sum_{j<i} \sigma_{i-j}w_{j} & (i \in \limitedlevels) 
 \end{LPNoCost}
&
\begin{LPNoCost}{constraints defining edge signature $e$}
 x_i + w_i & \le & r_i+\displaystyle\sum_{j<i} e_{ji} & (i \in \limitedlevels) 
\\
w_{ji} & = & \lceil e_{ji}/\sigma_{i-j} \rceil  & (i,j\in\limitedlevels, j<i)
\\
w_j & = & \displaystyle \max_{i>j}\, w_{ji}  & (j\in\limitedlevels)
\end{LPNoCost}
 \end{LPTAB}
\smallskip 

(Achieve this as follows.
For each $i\in\limitedlevels$, in increasing order,
choose $w_i$ maximally subject to the $i$th capacity constraint.
This assignment to $w$ will satisfy the capacity constraints (with $x$) if any assignment to $w$ can.)

2. In the edge-signature constraints on the right above,
$e_{ji}$ represents the number of edges from level $j$ to level $i>j$
and $w_{ji}$ represents the number of interior nodes in level $j$ with children in level $i>j$.
Initialize the edge signature $e$ and the $w_{ji}$'s
so that these constraints are met.
(To do this, take $e_{ji} = \sigma_{i-j}w_j$ and $w_{ji} = w_j$ for all $i$ and $j$.
Since the capacity constraints for $\MP$ are satisfied by $x$ and $w$,
by inspection, the edge-signature constraints for $e$ on the right above will also be satisfied.)

3. Next, {\em lower} $w$, $e$, and possibly $r$ so that all
of the edge-signature constraints above are tight.
(Achieve this by mimicking a leaves-to-root scan over the tree
that deletes ``unused'' interior nodes and edges, as follows.
For each $j\in\limitedlevels$, in {\em decreasing} order,
for each $i\in\limitedlevels$ with $i>j$,
lower $e_{ji}$ as much as possible
subject to the first edge-signature constraint for $i\in\limitedlevels$,
then update $w_{ji}$ and $w_i$.
Finally, if the first edge-signature constraint for some $i\in\limitedlevels$ is still loose,
it must be that $\sum_{j<i} e_{ji} = 0$,
so lower $r_i$ to $x_i+w_i$ to make the constraint tight.)

4.  In $F$, if for some edge $(a,b)$, $b$ is $a$'s only child, 
then call the node $a$ {\em useless.}  
(Contracting such edges would give a better code.)
Call all other nodes (including codeword nodes) {\em useful}.
For each $j$, count the number $u_{j}$ of useless nodes in level $j$ as follows.
For definiteness, order the level-$j$ nodes arbitrarily 
and assume that, for each $i,j\in\limitedlevels$ with $i>j$,
the nodes in level $j$ that have children in level $i$
are the {\em first} $w_{ji}$ interior nodes in level $j$,
and that all but the last of these $w_{ji}$ nodes has the maximum possible number
($\sigma_{j-i}$) of children in level $i$
(so that the last such node has $e_{ji}\bmod \sigma_{j-i}$ children in level $i$).
Then count the useless nodes in level $j$ as follows.
Let $i' = \arg\max_i w_{ji}$ and $i'' = \arg\max_{i\ne i'} w_{ji}$
be the two levels having the most and second-most children of nodes in level $j$.
(So $w_i = w_{ji'} \ge w_{ji''}$.)
If it happens that $\sigma_{j-i'}=1$,
then the last $w_j - w_{ji''}$ level-$j$ interior nodes have only one child, so $u_{j} = w_j - w_{ji''}$.
Otherwise ($\sigma_{j-i'} \ge 2$),
only the last level-$j$ interior node can have just one child
(because all others have $\sigma_{j-i'}$ edges to level $i'$).
The number of level-$i'$ children of that last node is $e_{ji'} \bmod \sigma_{j-i'}$.
If this quantity is 1 and $w_{ji''} < w_i$ (the node has no children in level $i''$),
then $u_{j} = 1$, and otherwise $u_{j} = 0$.

5.  Define $F'$ to be the sub-forest of $F$ induced by useful nodes and their children.
Explicitly construct $F'$, as follows.
For each level $j\in\limitedlevels$ in {\em decreasing} order, do the following.
Create the $x_j$ codeword nodes and the $w_j - u_j$ non-useless interior nodes.
Then, following the description of the edges in $F$ from Step 4 above,
for each $i>j$, add up to $e_{ji}$ edges greedily 
from each of the first $\min(w_{ji}, w_j-u_j)$ interior nodes
(adding at most $\sigma_{i-j}$ edges from each node)
to parentless nodes in level $i$ (giving those nodes parents).
If there are not enough parentless nodes in level $i$ to do this,
create new {\em childless} interior nodes in level $i$ as needed
(these new nodes are useless children of non-useless nodes;
in Step 6, below, they are the {\em stubs}).
Among all $x_j+w_j-u_j$ new nodes instantiated in level $j$,
designate as many as possible ($\min(x_j+w_j-u_j , r_j)$) as roots,
and designate the rest as (temporarily) parentless.
Non-root nodes might be left parentless
(these are nodes whose parents were useless in $F$; 
in Step 6, they are the {\em orphans}).

6.  Next consider the non-root parentless nodes in $F'$ (call these {\em orphans}),
and the (useless) childless interior nodes in $F'$ (call these {\em stubs}).
The nodes in $F-F'$ are interior nodes with one child whose parents also have one child,
so in $F$ the nodes in $F-F'$ form vertex-disjoint paths connecting each
orphan $d$ to a unique stub $A(d)$ (the child of $d$'s first non-redundant ancestor in $F$).
Thus, the number of orphans equals the number of stubs.
Make a list $a_1,a_2,\ldots,a_k$ of the stubs,
and a list $d_1,d_2,\ldots,d_k$ of the orphans
both ordered by increasing level (breaking ties arbitrarily).
Finally, modify $F'$ as follows.
For each pair of nodes $(a_j,d_j)$, identify $a_j$ and $d_j$
--- that is, make $d_j$ the child of $a_j$'s parent in place of $a_j$.  
The resulting forest is $F''$.

\correctness
Let $\code'$ be the monotone code with tree representation $F''$.
By construction, $\code'$ is prefix free, has codewords in $\universe$,
and has signature $x$.
To prove that $F"$ has cost no greater than the cost of $F$, 
we observe that each leaf node in $F$ has a corresponding leaf node in $F''$ and observe
that, in the last step of the construction, 
going from $F'$ to $F"$, cannot increase the level of any orphan $d_j$. 
Indeed, suppose for contradiction that the level of some $d_j$ in $F$
is strictly less than the level of its paired node $a_j$.
Thus, the $j$ stub nodes $A(d_1),A(d_2),\ldots,A(d_j)$
are in levels strictly less than the level of $a_j$.
Each of these $j$ nodes must precede $a_j$ in the ordering $a_1,a_2,\ldots,a_k$ of stub nodes,
but only $j-1$ nodes can do so.

\time
The time for constructing $x$, $w$, and $e$
is $O(|\limitedlevels|^2) = O_\eps(\log^2 n)$.
By inspection, the forest $F''$ can be constructed from $w$, $x$, and $e$ 
in time $O(|\limitedlevels|^2 + |F''|)$.
In $F''$ there are $n$ leaves and each interior node has at least two children,
so $|F''|\le 2n$.
\end{proof}

\section{Computing the signature of a near-optimal code when $\cost(1)\geq 3/\eps$}
\label{sec:general}

\newcommand{\HILR}{\HULR}
\newcommand{\LETTER}[1]{\ensuremath{\underline{#1}}}
\newcommand{\lifts}{\lift s}
\newcommand{\reservedWords}{\text{\bf FIXME}}

The preceding sections give a complete PTAS for instances of \HUL with $\cost(1)\le 3/\eps$.
In this section, the goal is to extend the PTAS to handle arbitrary letter costs.
Note that if the letter costs were fixed (not part of the input),
then for small (but still constant) $\eps$ it would be the case that $\cost(1)\le 3/\eps$,
so the PTAS in the preceding sections could be applied as it stands.
But since letter costs are part of the input, as we've defined \HUL,
we cannot assume $\cost(1)$ is constant;
we have to handle the case when $\cost(1)$ grows asymptotically.

Unfortunately,
the PTAS in the preceding sections makes fundamental use of the assumption that $\cost(1)= O_\eps(1)$.    
Indeed, that restriction is what ensures that the relaxation gap for $\threshold$-relax 
is $1+O(\eps)$ for some threshold $\threshold = O_\eps(1)$.
In turn, using a threshold $\threshold$
with value $O_\eps(1)$ is central to the polynomial running time.
This approach does not seem to extend to handle instances
in which the ratio $\cost(0)/\cost(1)$ is quite small (e.g., decreasing with $n$).
We need another approach for handling the case when $\cost(0)$ is quite small.

\subsection{Reducing to coarse letter costs}\label{subsec:coarse}
We start with a simple scaling and rounding step (a standard technique in PTAS's), 
to bring the letter costs into a restricted form that is easier to work with.
Ideally, we would like to make (i) all letter costs integers and (ii) $\cost(1) \le 3/\eps$,
for then the preceding PTAS would apply.
We almost achieve these two conditions,
failing only in that $\cost(0)$ may end up being non-integer.
More specifically, we scale and round the costs to make them {\em coarse}:
\begin{definition}
The letter costs are {\em coarse} if
\begin{itemize}

\item the second-cheapest letter cost, $\cost(1)$, is in the interval $[1/\eps,3/\eps]$; and

\item all letter costs are integers, except possibly $\cost(0)$, 
  which may instead be the reciprocal of an integer.
\end{itemize}
\end{definition}\smallskip

\noindent
Note well that {\em throughout this section $\cost(0)$ is not necessarily an integer}
--- it may instead be the reciprocal of an integer, i.e., $\cost(0)=1/N$ for some integer $N$.
All other letter costs are still integers.

Here are the specific scaling and rounding steps that we use to achieve coarse letter costs:
\begin{subroutine}[h]
\caption{--- {\em Coarsening the letter costs}
}\label{sub:coarsening}
\begin{algorithmic}[1]
\IF{$\cost(1)/\cost(0) \ge 1/\eps$}\label{step:scale1}
\STATE  Let $N$ be the maximum integer such that $\frac{\cost(1)}{N\,\cost(0)} \ge 1/\eps$.
\\Initialize $\cost'(\letter) = \frac{\cost(\letter)}{N\, \cost(0)}$ for $\ell\in\alphabet$.
\ELSE
\STATE \label{step:scale2}
Let $N$ be the minimum integer such that $\frac{N\,\cost(1)}{\cost(0)} \ge 1/\eps$.
\\Initialize $\cost'(\letter) = \frac{N\, \cost(\letter)}{\cost(0)}$ for $\ell\in\alphabet$.
\ENDIF
\STATE For each $\letter\in\alphabet$ except $\ell=0$, round $\cost'(\letter)$ to the integer $\lceil \cost'(\letter)\rceil$.
\label{step:round_cost}
\STATE Return $\cost'$.
\end{algorithmic}
\end{subroutine}

To conclude Section~\ref{subsec:coarse} we prove that the above procedure does
indeed produce coarse letter costs in linear time, and that any instance with arbitrary 
costs reduces (in an approximation-preserving way) to the same instance but
with coarsened costs.

\begin{sublemma}\label{lemma:coarse}
Let  $\cost':\alphabet\rightarrow\Rp$ be the costs output by
the coarsening subroutine
(given arbitrary letter costs $\cost:\alphabet\rightarrow\Rp$).
(i) The subroutine takes $O(n)$ time.
(ii) The costs $\cost'$ are coarse.
(iii) Any code that is near-optimal under $\cost'$ is also near-optimal under $\cost$.
\end{sublemma}

\begin{proof}
Part (i) is clear by inspection and the assumption that $|\alphabet|\le n$.  

(ii) If the condition in the ``if'' statement holds (that is, $\cost(1)/\cost(0)\ge 1/\eps$)
the scaling step makes $\cost'(0)$ $\big({=\frac{\cost(0)}{N\cost(0)} = 1/N}\big)$ the reciprocal of an integer.
Also, 
the scaling step bring $\cost'(1)$ into the interval $[1/\eps,2/\eps)$, because,
by the choice of $N$, 
\[\textstyle\frac{1}{2}\cost'(1) 
\le 
\frac N {N+1} \, \cost'(1) 
= \frac N {N+1} \, \frac {\cost(1)} {N \cost(0)} 
= \frac {\cost(1)} {(N+1) \cost(0)}
<
\frac{1}{\eps}
\le \frac {\cost(1)} {N \cost(0)}
= \cost'(1).\]

Alternatively, if the ``else'' clause is executed, the scaling step makes
$\cost'(0)$ an integer,
and brings $\cost'(1)$ into the interval $[1/\eps,2/\eps)$ because
$N\ge 2$ and
\[\textstyle\frac{1}{2}\cost'(1) 
\le 
\frac {N-1} {N} \, \cost'(1) 
= \frac {N-1} {N} \, \frac {N\cost(1)} {\cost(0)} 
= \frac {(N-1)\cost(1)} {\cost(0)}
<
\frac{1}{\eps}
\le \frac {N \cost(1)} {\cost(0)}
= \cost'(1).\]

In either case, the final rounding step (line~\ref{step:round_cost})
makes every $\cost'(\ell)$ (for $\ell\ge 1$) 
an integer.
The rounding step also leaves $\cost'(1)\le 3/\eps$,
because $\cost'(1)\le 2/\eps$ before rounding and $\lceil 2/\eps\rceil \le 3/\eps$ for $\eps\le 1$.

(iii) The scaling steps (lines~\ref{step:scale1}-\ref{step:scale2}) do not change the ratio of any two letter costs.
The rounding step changes the relative costs of any two letters by at most a factor of $1+\eps$,
because, before rounding, each rounded letter cost $\cost'(\ell)$ is at least $1/\eps$,
and so increases by at most a $1+\eps$ factor.
%
Thus,  any prefix-free code $\code$ is a near-optimal solution under $\cost'()$
iff it is a near-optimal solution under $\cost()$.
\end{proof}

\subsection{Reducing to coarse letter costs with $\cost(0)\geq 1$}
Appealing to Lemma~\ref{lemma:coarse}, 
we can now assume without loss of generality that the letter costs are coarse.
That is, we assume that $\cost(1) \in [1/\eps, 3/\eps]$,
and that all letter costs are integers except perhaps $\cost(0)$
which may instead be the reciprocal of an integer.

If it does happen that $\cost(0)$ is an integer, 
then the condition for the PTAS of the preceding sections is met:
all letter costs are integers and $\cost(1)\le 3/\eps$.
So in this case we can apply that PTAS directly to the instance.

So, assume that $\cost(0)$ is not an integer.
That is, $\cost(0)$ equals $1/N$ for some integer $N\ge 2$.

We now confront the core problem of this section:
how to deal with an instance in which $\cost(0)$ is very small in comparison to $\cost(1)$.
To handle such an instance, the basic idea is to reduce the problem to the case we've already solved.
In particular, we replace the given alphabet by a new alphabet $\lift\alphabet$, 
in which each new letter $\LETTER{s}$
represents some {\em string} $s$ over the original $\alphabet$.
This idea allows us to manipulate the letter costs: 
by choosing large enough strings $s$ to represent,
we can make sure no letter cost in $\lift\alphabet$ is too small.

\begin{figure}[t]
\centering
\includegraphics[width=\textwidth]{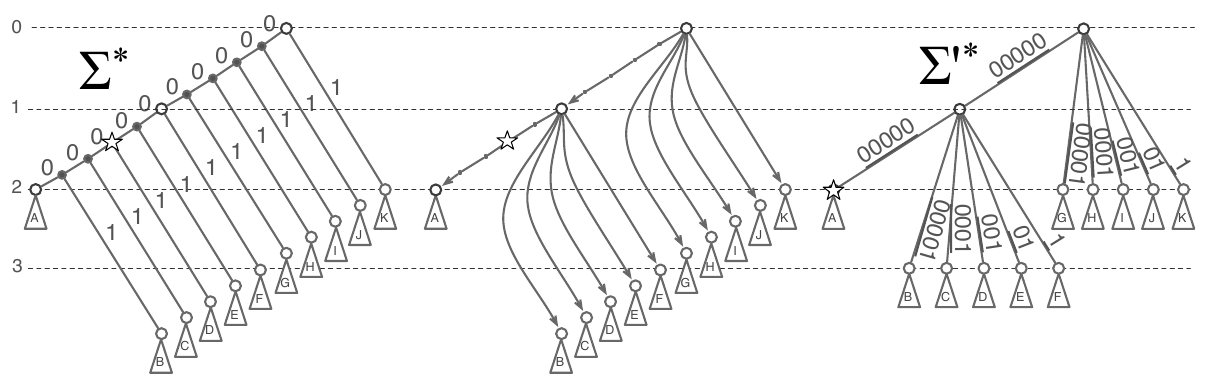}
\caption{\em Given $\alphabet=\{0,1\}$
with $\cost(0)=1/N=1/5$ 
and $\cost(1)=1/\eps = 2$,
the top of $\alphabet^*$ is on the left,
the top of $\lift\alphabet^*$ is on the right.
The ``chunk'' alphabet $\lift\alphabet$ has six letters, called ``chunks'':
the cheapest chunk, $\LETTER{00000}$, 
represents the string `$00000$' in $\alphabet^*$,
and costs 1.
The other five chunks
($\LETTER{1}$,
$\LETTER{01}$,
$\LETTER{001}$,
$\LETTER{0001}$,
and $\LETTER{00001}$)
represent, respectively, the strings
`${1}$',
`${01}$',
`${001}$',
`${0001}$',
and `${00001}$'.
Each costs $2=\cost(1)$.}
\label{fig:runt}
\hrulefill
\end{figure}

For intuition, consider an example with binary alphabet $\alphabet = \{0,1\}$.
Consider replacing this alphabet with an alphabet
$\lift\alphabet$ containing the six letters
$\LETTER{00000}$, $\LETTER{1}$, $\LETTER{01}$, $\LETTER{001}$, $\LETTER{0001}$, and $\LETTER{00001}$.
Call these letters {\em chunks}.
They represent, respectively, the four strings ${00000}$, ${1}$, ${01}$, ${001}$,
${0001}$, and ${00001}$ over $\alphabet$.
In this way, each string of chunks (i.e., string over $\lift\alphabet$)
represents a string over $\alphabet$ in a natural way,
For example,
the string `$\LETTER{1}~ \LETTER{00000} ~\LETTER{01}$' over $\lift\alphabet$
represents the string `$10000001$' over $\alphabet$.
See Fig.~\ref{fig:runt}.

For letter costs,
it would be natural to take $\cost(\LETTER{s})$ equal to the cost
of the string over $\alphabet$ that $\LETTER{s}$ represents.
For the example, if $\cost(0)$ is $1/5$,
it would be natural to take 
$\cost(\LETTER{00000}) = 1$,
$\cost(\LETTER{1}) = \cost(1)$,
$\cost(\LETTER{01}) = \frac{1}{5}+\cost(1)$,
$\cost(\LETTER{001}) = \frac{2}{5}+\cost(1)$, etc.
But, since our goal is to have all-integer letter costs, we instead round down the costs:
$\cost(\LETTER{00000}) = 1$,
$\cost(\LETTER{1}) = \cost(1)$,
$\cost(\LETTER{01}) = \cost(1)$,
$\cost(\LETTER{001}) = \cost(1)$, etc.
Because $\cost(1)\ge 1/\eps$,
rounding down doesn't alter the ``natural'' costs by more than a $1+\eps$ factor.

In general, for an arbitrary alphabet $\alphabet$, where, say $\cost(0)=1/N$,
here is how we construct $\lift\alphabet$:
\begin{subdefinition}[chunk alphabet]\label{defn:chunk}
Let {\em chunk} alphabet $\lift\alphabet$ contain the follow letters (called {\em chunks}):
one letter denoted $\LETTER{0^N}$ and, for each non-zero letter $\ell\in\alphabet$,
$N$ letters denoted $\LETTER{\ell},\LETTER{0\ell},\ldots,\LETTER{0^{N-1}\ell}$.
(Each underlined string $\LETTER{0^i\ell}$ denotes a single letter in $\lift\alphabet$.)
Give letter $\LETTER{0^N}$ cost 1 and give each letter $\LETTER{0^i\ell}$ cost equal to $\cost(\ell)$.
\end{subdefinition}\smallskip

For any string $\lift s$ over $\lift\alphabet$,
let $\unchunk(\lift s)$ denote the string over $\alphabet$ that $\lift s$ represents.
Say a string $s$ over $\alphabet$ is {\em chunkable}
if $s=\unchunk(\lift s)$ for some $\lift s$ over $\lift\alphabet$
(these are the strings over $\alphabet$ that can be cleanly broken into chunks).

Extending from strings to codes,
each code $\lift\code$ over $\lift\alphabet$ 
represents a code $\code$ over $\alphabet$ in a natural way, specifically
$\code_i = \unchunk(\lift\code_i)$.
Let $\unchunk(\lift\code)$ denote this code $\code$.
Say that a code $\code$ over $\alphabet$ is {\em chunkable}
if it can be obtained in this way (i.e., all its codewords are chunkable).

Thus, $\unchunk()$ gives a bijection between the strings over $\lift\alphabet$
and the chunkable strings over $\alphabet$.
Likewise, it gives a bijection between the codes over $\lift\alphabet$
and the chunkable codes over $\alphabet$.
On consideration, $\unchunk(\lift\code)$ will be prefix-free if and only if $\lift\code$ is prefix-free.
Thus, this bijection preserves prefix-free-ness and (approximate) cost.

\paragraph{First attempt at PTAS via reduction}
The general scheme will be something like the following:
\smallskip
\begin{center}\noindent
\parbox{0.95\textwidth}{
\hrulefill

(1) Given $\alphabet$, construct the chunk alphabet $\lift\alphabet$.

(2) Find a near-optimal prefix-free code $\lift\code$ over $\lift\alphabet$
using PTAS for $\cost(1)\le 3/\eps$.

(3) Return the prefix-free code $\code=\unchunk(\lift\code)$ that $\lift\code$ represents.

\hrulefill
}
\end{center}

\smallskip
The main flaw in this reduction is the following: 
not all strings over $\alphabet$ can be broken into chunks from $\lift\alphabet$.
In particular, the codewords in the optimal code $\code^\smallopt$ over $\alphabet$ 
might not be chunkable.
Thus, even if $\lift\code$ is near-optimal over $\lift\alphabet$,
a-priori, it may happen that $\unchunk(\lift\code)$ is far from optimal over $\alphabet$.

The main technical challenge in this section is to understand this flaw and work around it.
To understand the flaw in detail, recall that the codes over $\lift\alphabet$ correspond,
via the bijection $\unchunk()$,
to the chunkable codes over $\alphabet$,
and this bijection preserves prefix-free-ness and (approximately) cost.

Because of this bijection,
the reduction proposed above (after Defn.~\ref{defn:chunk})
will work if and only if the optimal prefix-free code $\code^\smallopt$ over $\alphabet$
is has approximately the same cost as the optimal {\em chunkable} prefix-free code over $\alphabet$
(since the latter code has approximately the same cost as the optimal prefix-free code over $\lift\alphabet$).
So, is there always a chunkable prefix-free code 
whose cost is near that of the optimal prefix-free code $\code^\smallopt$?

Let's consider
which strings over $\alphabet$ are chunkable (that is, can be broken into chunks from $\lift\alphabet$).
On consideration,\footnote
{A string with this property can be broken into chunks as follows:
first break the string after each occurrence of each non-zero letter,
leaving pieces of the form $0^i\ell$ for some $i$, plus a final piece of the form $0^{iN}$ for some $i$;
then, within each such piece, break the piece after every $N$'th 0.}
a necessary and sufficient condition for a string $s$ over $\alphabet$
to be chunkable is that the number of `0's at the end of $s$ should be a multiple of $N$.
Thus, a given code $\code$ over $\alphabet$ is chunkable
if and only if all of its codewords end nicely in that way.
Define $\pad(\code)$ to be the code over $\alphabet$ obtained by
padding each codeword in $\code$ with just enough `0's 
so that the number of `0's at the end of the codeword is a multiple of $N$.

Then $\pad(\code^\smallopt)$ is a prefix-free, chunkable code over $\alphabet$.
But how much can padding increase the cost of $\code^{\smallopt}$?
Padding a codeword adds at most $N-1$ `0's to the codeword.
This increases each codeword cost by at most $(N-1)\cost(0) = (N-1)/N < 1$.

Is this significant?
That is, can it increase the cost of the codeword by more than a $1+\eps$ factor?
In order for this to happen, the codeword must have cost less than $1/\eps$.
Call any such codeword (of cost less than $1/\eps$) a {\em runt}.
Recalling that $\cost(\ell)\ge 1/\eps$ for every letter $\ell\in\alphabet-\{0\}$,
for a codeword in $\code^{\smallopt}$ to be a runt
it must consist only of `0's.
In any prefix-free code, there is either one runt or none,
and the only codeword that can be the runt is the cheapest one, $\code^\smallopt_1$.

In sum, the reduction above fails, but just barely,
and the reason that it fails is because padding the runt 
can, in the worst case, increase the cost of the code by too much.

\paragraph{Second attempt}
To work around this issue, we handle the runt differently: we use exhaustive search
to remove it from the problem, then solve the remaining runt-free problem as described above.

More specifically, we consider all possibilities for the runt in the optimal code:
either the optimal code has no runt (in which case the reduction in the first attempt above works),
or the optimal code has a runt of the form $0^q$ for some $q\le n$ such that $\cost(0^q)< 1/\eps$.
For each possible choice $0^q$ for $\code_1$,
we compute a near-optimal choice for the $n-1$ remaining codewords 
$\code_2, \code_3,\ldots, \code_{n}$ given that $\code_1=0^q$.
We then return the best code found in this way.

How do we find a near-optimal choice 
for the $n-1$ {\em remaining} codewords given a particular choice $0^q$ for $\code_1$?
This problem can be stated precisely as:
\begin{equation}\label{eq:runt1}
  \left|~~
\parbox{0.85\linewidth}{\small
 Find a near-optimal prefix-free code of $n-1$ codewords over alphabet $\alphabet$, 
  \\ for probabilities $\prob' = \langle\prob_2,\prob_3,\ldots,\prob_n\rangle/(1-p_1)$,
  \\ {\em from the universe $\universe_q$ of strings that do not have $0^q$ as a prefix}.
}\right.
\end{equation}

Since padding any non-runt codeword to make it chunkable increases its cost by at most a $1+\eps$ factor
and maintains prefix-free-ness,
the problem above reduces in an approximation-preserving way to the following one:
\begin{equation}\label{eq:runt2}
  \left|~~
\parbox{0.85\linewidth}{\small
 Find a near-optimal prefix-free code of $n-1$ codewords over alphabet $\alphabet$, 
  \\ for probabilities $\prob' = \langle\prob_2,\prob_3,\ldots,\prob_n\rangle/(1-p_1)$,
  \\ from the universe $\hat\universe_q$ of {\em chunkable} strings that do not have $0^q$ as a prefix.
}\right.
\end{equation}

Since the chunkable strings over $\alphabet$ correspond via the bijection $\unchunk()$
to the strings over chunk alphabet $\lift\alphabet$, and this bijection preserves 
prefix-free-ness and approximate cost,
the problem above in turn reduces in an approximation-preserving way to the following problem:
\begin{equation}\label{eq:runt3}
  \left|~~
\parbox{0.85\linewidth}{\small
 Find a near-optimal prefix-free set of $n-1$ codewords over {\em chunk alphabet $\lift\alphabet$}, 
  \\ for probabilities $\prob' = \langle\prob_2,\prob_3,\ldots,\prob_n\rangle/(1-p_1)$,
  \\ from universe $\lift\universe_q$ of strings {\em $s$ such that $\unchunk(s)$ 
    does not have $0^q$ as a prefix}.
}\right.
\end{equation}

Note that the chunk alphabet $\lift\alphabet$ in the latter problem~(\ref{eq:runt3}) 
has integer letter costs,
and the second cheapest letter cost is $\cost(\LETTER{1}) = \cost(1)$, which is in $[1/\eps,3/\eps]$.
These letter costs are appropriate for the PTAS from the preceding sections.
We solve problem~(\ref{eq:runt3}) using that PTAS.

To do so we have to limit the codeword universe $\universe =\lift\universe_q$
to those ``strings $s$ such that $\unchunk(s)$ does not have $0^q$ as a prefix.''
The basic idea is to choose an appropriate root set $\lift\roots_q$ for $\lift\universe_q$.
For intuition, consider an example
with binary alphabet $\alphabet=\{0,1\}$, with $\cost(0)=1/5$ and $\cost(1)=2$.
The strings over $\alphabet$ are shown to the left;
the strings over the chunked alphabet $\lift\alphabet$ are shown to the right.
A potential runt $0^7$ is marked with $\star$.
The strings having $0^7$ as a prefix (on the left)
and the corresponding strings over $\lift\alphabet$ ($s$ such that $\unchunk(s)$
has $0^7$ as a prefix, on the right) are gray:
\begin{figure}[h]\centering
\includegraphics[width=0.8\textwidth]{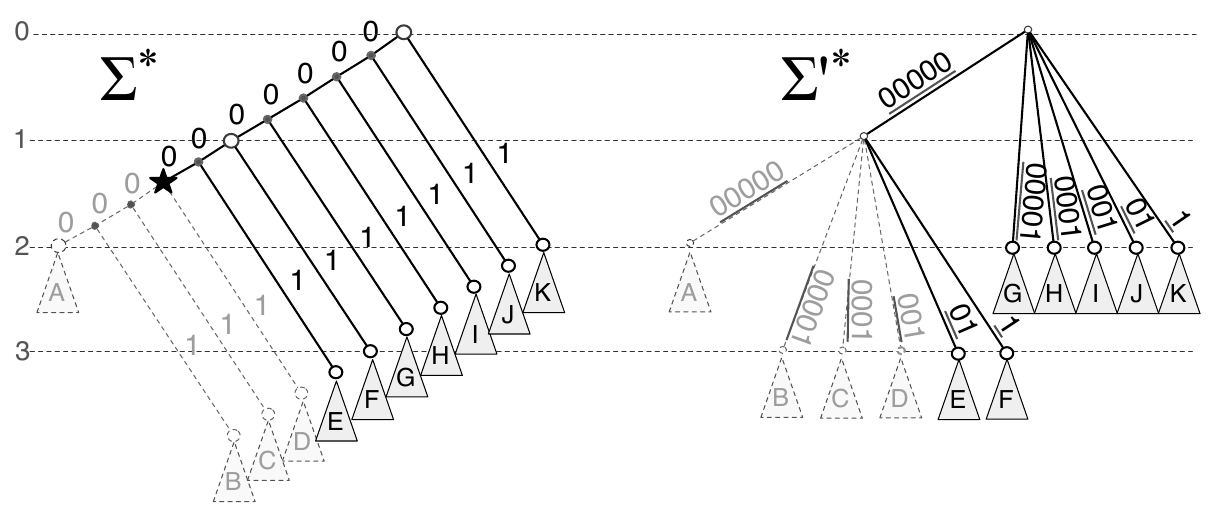}
\end{figure}

The remaining (allowed) strings are those in the subtrees marked $E,F,\ldots,K$
(on both the left and the right).
The roots of these subtrees are the roots of $\lift\universe_7$.


In general, given any alphabet $\alphabet$ where $\cost(0)=1/N$ for some integer $N$,
and given an arbitrary runt $0^q$,
we compute the root set $\lift\roots_q$ for the desired universe $\lift\universe_q$ as follows.

Let $\chunk()$ denote the functional inverse of $\unchunk()$:
if string $s$ is chunkable, then $\chunk(s)$ 
is the string $\lift s$ over $\lift\alphabet$ such that $\unchunk(\lift s) = s$;
likewise, if code $\code$ is chunkable, then $\chunk(\code)$ 
is the code $\lift\code$ over $\lift\alphabet$ such that $\unchunk(\lift\code) = \code$.

The universe $\lift\universe_q$ should contain those strings $\lift s$
such that $\unchunk(\lift s)$ does not have $0^q$ as a prefix.
The chunkable strings over $\alphabet$ that do not have $0^q$ as a prefix
are those that start with a prefix of the form $0^i\ell$ where $i<q$ and $\ell\in\alphabet - \{0\}$.
Each such string  $0^i\ell$ is itself chunkable (as it ends in a letter other than `0').
Thus, $\unchunk(\lift s)$ does not have $0^q$ as a prefix
iff $\lift s$ starts with a prefix of the form $\chunk(0^i\ell)$ where $i<q$ and $\ell\in\alphabet - \{0\}$.
That is, the universe $\lift\universe_q$ has root set
$\lift\roots_q = \{\chunk(0^i\ell) ~:~ i<q, \ell\in\alphabet - \{0\}\}$.

Thus, we can reformulate problem~(\ref{eq:runt3}) with an explicit root set as
\begin{equation}\label{eq:runt4}
\left|~~
\parbox{0.85\linewidth}{\small
  Find a near-optimal prefix-free set of $n-1$ codewords over chunk alphabet $\lift\alphabet$, 
  \\ for probabilities $\prob' = \langle\prob_2,\prob_3,\ldots,\prob_n\rangle/(1-p_1)$,
  \\ from the universe $\lift\universe_q$ with root set
  $\lift\roots_q = \{\chunk(0^i\ell) ~:~ i<q, \ell\in\alphabet - \{0\}\}$.
}\right.
\end{equation}
We solve this problem using the PTAS from the preceding sections.

\newcommand{\subcode}{\calY}




Next is a precise summary of the entire reduction.

For efficiency, instead of considering all possible choices $0^q$ for the root 
(for all $q<n$ such that $\cost(0^q)< 1/\eps$),
we further restrict $q$ to be near a power of $1+\eps$.
This is okay because in any prefix-free code
the runt $0^q$ can be padded with $O(\eps q)$ `0's
to convert it to this form,
without increasing the cost by more than a $1+\eps$ factor.
(This reduces the number of possibilities for the runt
from $n$ to $O_\eps(\log n)$.)

\begin{subdefinition}[reduction]\label{def:reduction}
  {\bf Forward direction:}
    Given a \HUL\ instance $\instance=(\prob,\alphabet,\cost)$,
    the forward direction of the reduction
    produces a set of instances
    $\{\lift\instance_{0}\} \cup \{\lift\instance_{q} ~|~ q\in Q\}$
    over alphabet $\lift\alphabet$,
    where $Q=\{\lceil \min(n, N/\eps)/(1+\eps)^j\rceil ~|~ j\in \Nz\}$
    (one instance for each choice of runt in \opt).

    Instance $\lift\instance_{0}$ 
    (for the case of no runt in \opt)
    is
    $(\prob,\lift\alphabet,\cost,\lift\roots_{0})$
    with chunked alphabet $\lift\alphabet$ 
    and universe $(\lift\alphabet)^*$
    (with root set $\lift\roots_{0}$ containing just the empty string).

    For each $q\in Q$, 
    instance $\lift\instance_q$ 
    (for the case of runt $0^q$ in \opt)
    is $(\lift \prob, \lift\alphabet, \cost, \lift\roots_q)$
    where
    $\lift \prob = \langle\prob_2, \prob_3, \ldots, \prob_n\rangle/(1-\prob_1)$
    and universe $\lift\universe_q$
    contains the string $s$ over $\lift\alphabet$ such that $\unchunk(s)$ 
    doesn't have $0^q$ as a prefix
    (root set $\lift\roots_q = \{\chunk(0^i\ell) ~:~ i<q, \ell\in\alphabet - \{0\}\}$).
    \smallskip

  \noindent{\bf Backward direction:}
  Given any near-optimal prefix-free code $\subcode^{0}$ for $\lift\instance_{0}$,
  and near-optimal prefix-free codes $\subcode^q$ for each $\lift\instance_q$,
  the reverse direction of the reduction produces a near-optimal code $\code^{\min}$
  for the original instance $\instance$ as follows:

  Let $\code^{\min}$
  by a code of near-minimum cost among 
  the codes
  $\unchunk(\subcode^{0})$,
  and $\{0^q\} \cup \unchunk(\subcode^q)$
  for $q\in Q$.
  Return $\code^{\min}$.
\end{subdefinition}\smallskip
\smallskip

By the preceding discussion, the reduction above is correct:

\begin{sublemma}[correctness]\label{lemma:reduction}
Assuming the codes $\subcode^q$ for $q\in\{0\}\cup Q$ are near-optimal prefix-free codes
for their respective instances,
the code $\code^{\min}$ returned by the reduction above
is a near-optimal prefix-free code for $\instance$.
\end{sublemma}

\begin{proof}
  By construction, all of the codes
  $\unchunk(\subcode^{0})$,
  and $\{0^q\} \cup \unchunk(\subcode^q)$
  for $q\in Q$,
  are prefix-free codes over $\alphabet$.

  To see that at least one of these codes is near-optimal,
  let $\code^\smallopt$ be an optimal prefix-free code over $\alphabet$.
  In the case that $\code^{\smallopt}$ has no runt, 
  the code $\chunk(\pad(\code^\smallopt))$ 
  for instance $\lift\instance_0$
  has approximately the same cost as $\code^\smallopt$,
  so the code $\subcode^0$ for $\lift\instance_0$
  also has approximately the same cost as $\code^\smallopt$,
  and thus so does $\unchunk(\subcode^0)$.
  
  Otherwise code $\code^\smallopt$ has some runt $0^q$ with $q\le \min(n, N/\eps)$.
  Padding the root to $0^{q'}$ for $q'\in Q$ 
  gives a prefix-free code $\code$ over $\alphabet$ of approximately the same cost.
  By construction, for the near-optimal solution $\subcode^{q'}$ to instance $\lift\instance_{q'}$,
  the codewords in $\unchunk(\subcode^{q'})$ are a near-optimal choice for the non-runt codewords
  for any code over $\alphabet$ with runt $0^{q'}$.
  Thus, the cost of the prefix-free code $\{0^{q'}\} \cup \unchunk(\subcode^{q'})$
  is approximately the same as $\cost(\code)$,
  which is approximately the same as $\cost(\code^{\smallopt})$.
\end{proof}

\subsection{Proof of Theorem~\ref{thm:main}}
The full PTAS implements the reduction in Defn.~\ref{def:reduction}.
That is, it uses the PTAS from the preceding section to approximately 
solve the instances $\{\lift\instance_q\}_q$ produced by the forward direction of the reduction,
then computes and returns $\code^{\min}$ following the backward direction of the reduction.
By Lemma~\ref{lemma:reduction}, 
this gives a near-optimal prefix-free code for the given instance.
Below is an outline of the steps needed to achieve running time $O(n) + O_\eps(\log^3 n)$.

\paragraph{Step 1 (forward direction --- computing and solving the instances)} 
For each of the $O_\eps(\log n)$ instances $\{\lift\instance_q\}_q$,
the PTAS first computes the signature and approximate cost (not the code tree) 
of the respective solutions $\{\lift\subcode_q\}_q$,

By Thm.~\ref{thm:cost1sig}  (Section~\ref{subsec:cost1sig proof}),
for each instance $\lift\instance_q$, the signature
and approximate cost of the solution $\subcode^q$ can be computed in $O_\eps(\log^2 n)$ time
given appropriate precomputed inputs.  Here is a restatement of that theorem:

\thmcostsig

To solve the instances this way, we need to precompute three things for each instance:
the cumulative probability distribution,
the signature of the chunked alphabet $\lift\alphabet$,
and the signature of the root set.

Regarding the cumulative probability distributions, in fact there are only two distinct 
distributions used by the instances: $\prob$ for $\lift\instance_0$,
and $\lift\prob$ for the remaining instances.  So the necessary cumulative distributions
for all instances can be computed in $O(n)$ time.

Regarding the signature $\lift\sigma$ of $\lift\alphabet$,
it can be computed as follows.
First, compute the signature $\sigma$ of $\alphabet-\{0\}$ in $O(n)$ time.
Then, according to the definition 
$\lift\alphabet = \{\LETTER{0^N}\} \cup \{\LETTER{0^i\ell}~:~ i < N, \ell\in\alphabet-\{0\}\}$,
take $\lift\sigma_1 = 1$ and,
for $j$ such that $\sigma_j > 0$,
take $\lift\sigma_j = N \sigma_j$.
This takes $O(n)$ time since $|\alphabet|\le n$.

Next consider how to compute the root-set signatures.
For $\lift\instance_0$, the root set is trivial.
For each of the remaining $O_\eps(\log n)$ instances $\lift\instance_q$,
the PTAS computes the signature of the root set in $O_\eps(\log n)$ time using the following lemma:
\begin{sublemma}\label{lemma:root sig}
Given the signature $\sigma$ of $\alphabet-\{0\}$,
the signature $r^q$ of the root set of universe $\lift\universe_q$ for $\lift\instance_q$
(restricted to the set $\limitedlevels$ of possible levels, per Observation~\ref{obs:limitedlevels})
can be computed in time $O_\eps(\log n)$.
\end{sublemma}
\smallskip

\begin{proof}
  The root set for instance $\lift\instance_q$ is
  $\lift\roots_q = \{\chunk(0^j\ell) ~:~\ell\in\alphabet - \{0\}, 0\le j<q\}$.

  The associated multiset of costs is
  $\{\cost(\chunk(0^j\ell)) ~:~ \ell\in\alphabet - \{0\}, 0\le j<q\}$.

  Expressing the costs explicitly, this is
  $\{\lfloor j/N\rfloor + \cost(\ell) ~:~ \ell\in\alphabet - \{0\}, 0\le j<q\}$.
  \smallskip

  In this multiset, by calculation, each fixed $\ell\in\alphabet-\{0\}$
  contributes $N$ copies of $a+\cost(\ell)$ for each non-negative integer $a < \lfloor q/N\rfloor$,
  and $q \bmod N$ copies of $\lfloor q/N\rfloor+\cost(\ell)$.
  Thus, the multiset can be expressed as 
  \begin{eqnarray*}
     N& \times & \{a+\cost(\ell) ~:~ \ell\in\alphabet - \{0\}, 0\le a<\lfloor q/N\rfloor\}\\
     \bigcup~~~ (q\bmod N) &\times& \{a+\cost(\ell) ~:~ \ell\in\alphabet - \{0\}, a = \lfloor q/N\rfloor\}.
  \end{eqnarray*}
  Introducing variable  $i=\cost(\ell)+a$ to eliminate $a$, and rearranging the inequalities, this is
  \begin{eqnarray*}
    N& \times & \{i ~:~\ell\in\alphabet - \{0\}, i-\lfloor q/N\rfloor <\cost(\ell) \le i\}\\
    \bigcup~~~ (q\bmod N) &\times& \{i ~:~ \ell\in\alphabet - \{0\}, \cost(\ell) = i - \lfloor q/N\rfloor\}.
  \end{eqnarray*}
  Thus, introducing variable $j=\cost(\ell)$ 
  and recalling that $\sigma_j$ is the number of cost-$j$ letters in $\alphabet-\{0\}$, 
  a given $i\in\limitedlevels$ occurs with multiplicity
  \[
  r^q_i ~=~~ N \times \sum_{j=i-\lfloor q/N\rfloor+1}^{i} \sigma_j
  ~~~~+~~~ (q\bmod N) \times \sigma_{i-\lfloor q/N\rfloor}.
  \]
  Since by assumption $0^q$ is a runt, $\cost(0^q)< 1/\eps$, so $q/N<1/\eps$.
  Thus, the sum above has at most $1/\eps$ terms,
  and the value of $r^q_i$ for a given $i$ and $q$ can be calculated in $O(1/\eps)$ time.

  To finish, we observe that the set $\limitedlevels$ of possible levels 
  for the instance $\lift\instance_q$ can be computed as follows.
  Per Defn.~\ref{def:limitedlevels} (Section~\ref{subsec:limitedlevels}),
  the set is
  \([0,2\threshold+3\delta]
  ~\cup~ [i_{\lift\roots_q}, i_{\lift\roots_q} + 3\delta] 
  ~\cup~ [i_{\lift\alphabet}, i_{\lift\alphabet} + \threshold + 3\delta]
  \).

  The values of $\threshold$ and $\delta$ 
  (resp., $\lceil\log_2[\cost(\LETTER{1})/\eps] \cost(\LETTER{1})/\eps\rceil$
  and $\cost(\LETTER{1})\lceil\log_2 n\rceil$) are easy to calculate.
  
  By definition, $i_{\lift\alphabet}$
  is the minimum cost of any letter in $\lift\alphabet$ of cost at least $\threshold$.
  It can be calculated (just once) in $O(\log n)$ time by binary search over $\alphabet$.

  By definition, $i_{\lift\roots_q}$
  is the minimum cost of any root in $\lift\roots_q$ of cost at least $\threshold$.
  Reinspecting the calculation of $r^q_i$ above,
  $i_{\lift\roots_q}$ is the minimum value of the form $a+\cost(\ell)$ exceeding $\threshold-1$,
  for any $\ell\in\alphabet-\{0\}$ and integer $a\in [0, q/N]$.
  This value can be found in binary search over $\alphabet$ in $O(\log n)$ time.
  
  \smallskip
  Once $\limitedlevels$ is computed for $\lift\instance_q$,
  each coordinate of the signature $r^q$ of the root set $\lift\roots_q$ above (restricted to $\limitedlevels$)
  can be calculated in $O_\eps(1)$ time.
  Since $|\limitedlevels|=O_\eps(\log n)$, the total time is $O_\eps(\log n)$.
\end{proof}
\smallskip

In sum, the PTAS pre-computes the necessary inputs for all instances $\{\lift\instance_q\}_q$
of the reduction, taking $O_\eps(\log n)$ time for each of the $O_\eps(\log n)$ instances.
It then applies Thm.~\ref{thm:cost1sig} to solve these instances.
Specifically, in $O(n) + O_\eps(\log^3 n)$ total time,
it computes the signature and approximate cost 
of a near-optimal prefix-free code $\subcode^q$ for every instance $\lift\instance_q$.

\smallskip
\paragraph{Step 2 (backward direction --- building the near-optimal code tree)}
The backward direction of the reduction must
return a near-minimum-cost code $\code^{\min}$ among
the following candidate codes:
$\unchunk(\subcode^0)$, and $\{0^q\}\cup \unchunk(\subcode^q)$ for $q\in Q$.

At this point, the PTAS has only the signatures and approximate costs
of the various codes $\{\subcode^q\}_q$.  But this is enough information
to determine which of the candidate codes above have near-minimum cost.
In particular, unchunking a code approximately preserves its cost,
so the PTAS knows the approximate costs of each code $\unchunk(\subcode^q)$.
Then, from the approximate cost of $\unchunk(\subcode^q)$,
the approximate cost of $\{0^q\}\cup \unchunk(\subcode^q)$ is easily calculated.
(Recall that each code $\unchunk(\subcode^q)$
is for probabilities $\lift\prob = \langle\prob_2, \prob_3, \ldots, \prob_n\rangle/(1-\prob_1)$
and has non-runt codewords that don't have $0^q$ as a prefix.
By calculation,
adding codeword $0^q$ to the code gives a code for $\prob$ of cost
$\prob_1 \cost(0^q) + (1-\prob_1) \cost(\unchunk(\subcode^q))$.)
In this way, the PTAS chooses the index $q$ of the best candidate code.
The PTAS retains the signature $x$ of the corresponding code $\subcode^q$ over $\lift\alphabet$.

One more step remains: to compute the tree representation $\tree$
of the chosen candidate code $\code^q$ (i.e., $\unchunk(\subcode^0)$ if $q=0$,
or $\{0^q\}\cup \unchunk(\subcode^q)$ if $q>0$).

Recall that, by Thm.~\ref{thm:cost1tree},
for alphabets with integer letter costs and $\cost(1)\le 3/\eps$,
given a signature $x$ for a prefix-free code,
one can compute a corresponding tree representation $\lift\forest$ in $O(n)+O_\eps(\log^2 n)$ time.
This theorem doesn't solve our problem directly
for two reasons: (1) the signature $x$ that we have is 
for the code $\subcode^q$ over chunk alphabet $\lift\alphabet$,
not for the final code $\code^q$ over $\alphabet$;
(2) more fundamentally, because $\cost(0) = 1/N < 1$ for $\alphabet$,
the concept of signature is not particularly useful when working over $\alphabet$.

Instead, to compute the tree $\tree$ for $\code^q$,
the PTAS uses Thm.~\ref{thm:cost1tree} to first compute
the tree representation $\lift\forest$ of the prefix-free $\subcode^q$ over $\lift\alphabet$.
This takes $O(n)+O_\eps(\log^2 n)$ time.
The PTAS will then convert this tree representation $\lift\forest$ for $\subcode^q$ 
directly into a tree $\tree$ for $\code^q$,
using the following lemma:

\begin{sublemma}\label{lemma:lift tree}
(i) Given the tree $\lift\forest$ for a code $\subcode^0$ for $\lift\instance_0$,
the tree $\tree$ for the corresponding code $\code^0 =\unchunk(\subcode^0)$
(or one at least as good)
for $\instance$ can be constructed in $O(|\lift\forest|)$ time.

(ii) Given the forest $\lift\forest$ for a code $\subcode^q$ for $\lift\instance_q$,
the tree $\tree$ for the corresponding code $\code^q = \{0^q\}\cup\unchunk(\subcode^q)$ 
(or one at least as good) 
for $\instance$ can be constructed in $O(|\lift\forest|)$ time.
\end{sublemma}

\begin{proof}
\noindent
{\em Part (i).}
In this case ($q=0$), $\subcode^0$ has $n$ codewords and $\lift\forest$ has only a single root node.
$\lift\forest$ is the tree representation of $\subcode^0$ over $\lift\alphabet$,
and we want to compute the tree representation of $\code^0 = \unchunk(\subcode^0)$ over $\alphabet$.
Note that the $\unchunk()$ function simply breaks each chunk $\LETTER{0^N}$
or $\LETTER{0^i\ell}$ into its individual letters over $\alphabet$.

In the tree representation $\lift\forest$ of $\subcode^0$ over $\lift\alphabet$,
each edge (such as $\LETTER{0001}$) represents a chunk.
To ``unchunk'' the tree, we replace each such edge 
by a path (such as $0\rightarrow 0\rightarrow 0 \rightarrow 1$),
adding intermediate nodes as necessary.
This can be accomplished by applying a local transformation 
at each interior node $\lift u$ of $\lift\forest$, as illustrated (from right to left) here:
\begin{center}
\includegraphics[width=0.6\textwidth]{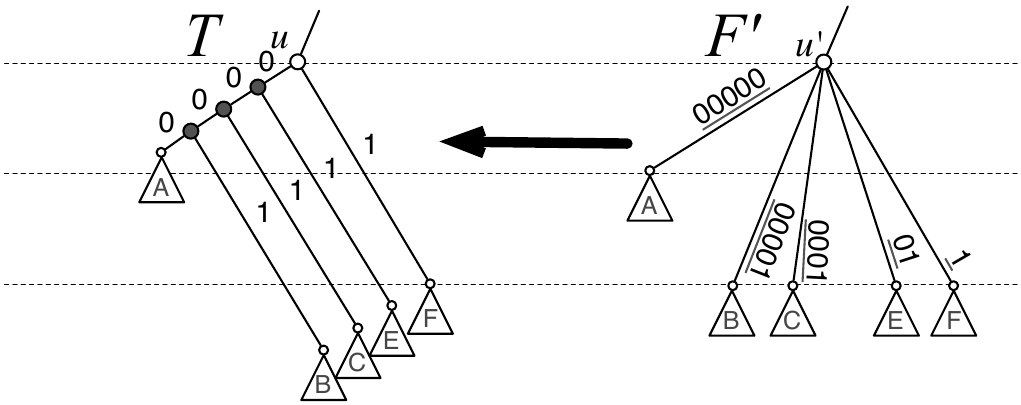}
\end{center}
Roughly,
for each edge labeled $\LETTER{a_1a_2\ldots a_k}$ on the right,
there is a corresponding path $a_1\rightarrow a_2 \rightarrow \cdots \rightarrow a_k$ on the left.
But
for efficiency, in fact we do something slightly different.
In general, in the tree $\lift\forest$, 
only some of the possible edges might be present.
(For example, there is no edge labeled $\LETTER{001}$ out of $\lift u$ on the right.)
When there are vacancies such as this, we first preprocess the node,
replacing edges in $\lift\tree$ by cheaper edges if possible.
In general, if the node $\lift u$ has some $d$ children,
we preprocess the node to make sure those $d$ children
use the $d$ cheapest possible outgoing edges in $\lift\alphabet$.
In this way we avoid constructing overly large trees.

The general construction is as follows.
For each interior node $\lift u$ in $\lift\forest$,
let $\lift v_0$ be the child along the edge labeled $\LETTER{0^N}$, if any.
Let $\lift v_1, \ldots, \lift v_d$ be the remaining children. 
Replace the edges to these latter $d$ children by a subtree $t(d)$ with $d$ leaves,
where $t(d)$ is the tree representation for the $d$ cheapest strings in
$\{b~|~\LETTER b \in\lift\alphabet, b \ne 0^N\}$.
Next, identify each child $\lift v_i$ for $i\ge 1$ with the $i$th cheapest leaf in $t(d)$.
Then make $\lift v_0$ the 0-child of the node at the end of the left spine of $t(d)$.
Doing this for all interior nodes gives $\tree$.

\smallskip\noindent{\em Part (ii).}
In this case the forest $\lift\forest$ is a collection of trees, 
each with its own root in the root set $\lift\roots_q$ of $\lift\universe_q$.
Let $d'$ be the number of roots.
Perform the transformation described in Part (i) separately for each tree in the collection.
Finally, glue the $d'$ trees together into a single tree $\tree$ as follows:
start with a tree whose $d'$ leaves are the $d'$ cheapest roots in the root set,
then, for $j=1,2,\ldots,d'$, identify the $j$th of these leaves with the root
of the $j$th (modified) tree in the collection.
Finally, add a leaf $0^{q'}$ where $q'\le q$ is the minimum
such that $0^{q'}$ is not already an interior node in $\tree$.

\correctness
By inspection of the construction, each leaf node $\lift v$ in $\lift\forest$
becomes a leaf in $\tree$ whose cost is at most $1+\eps$ 
times the cost of the string over $\alphabet$
that the string of $\lift v$ originally represented.
If $q>0$, the runt in $\tree$ has at most $q$ zeros,
so has cost at most $\cost(0^q)$.

\time
Assuming that each interior node $\lift u$  in $\lift\forest$ comes with a list of 
the edges to its children ordered by increasing cost,
the local transformation at each node $\lift u$ can be done in
time proportional to its degree $d$.
Also, gluing together the roots takes time proportional
to the number of roots, since in the resulting tree each
interior node has degree at least two
(recall that the roots unchunk to strings of the form $0^j\ell$ for $j<d$,
which hang consecutively off the left spine of $\tree$).
Thus, the entire transformation can be done in time proportional
to the size of $\lift\forest$.
\end{proof}

Since the trees produced via Thm.~\ref{thm:cost1tree} have size $O(n)$,
the time the PTAS takes to construct the tree $\tree$
for the near-optimal code $\code^q$ 
via Lemma~\ref{lemma:lift tree}
is $O(n)$.

This completes the PTAS and the Proof of Thm.~\ref{thm:main}.
\hfill\endproof

\section{Remarks}

\paragraph{More precise time bound}
The proof of Thm.~\ref{thm:main} shows that the PTAS runs in $O(n) + O_\eps(\log^3 n)$ time.
We note without proof that the time is 
\[O(n) + \exp\Big(O\Big(\frac 1 {\eps^3} \log^2 \frac 1 \eps\Big)\Big)\log^3 n.\]
Here is a sketch of the reasoning.
By careful inspection of the proof of Thm.~\ref{thm:main},
the time is proportional to $(\threshold+1)^\gamma(|\limitedlevels|^2 + \gamma) \,  |Q|$.
Plugging in $\threshold=O(\eps^{-2}\log \eps^{-1})$
(Lemma~\ref{lemma:relax}),
$\gamma = O(\threshold/\eps)$
(Defn.~\ref{def:grouping}),
$|\limitedlevels| = O(\threshold+\eps^{-1}\log n)$
(Defn.~\ref{def:limitedlevels}), 
and $|Q| = O(\eps^{-1}\log n)$
(Defn.~\ref{def:reduction})
gives the claim.
(Slightly better bounds can be shown with more careful arguments,
including coarsening the letter costs to ceilings of powers of $(1+\eps)$
to reduce the number of distinct letter costs.)

\paragraph{Practical considerations}
The exhaustive search outlined in
Section~\ref{subsec:cost1sig proof}
is the bottleneck of the computation.
In practice, this search can be pruned
and restricted to {\em monotone} group-to-level assignments.
Or, it may be faster to use a mixed integer-linear program solver
to solve the underlying program.
In this case, the alternate mixed program in Fig.~\ref{fig:alternative} may be easier to solve than $\MPG$, 
as it integer (in fact 0/1) variables only for the probabilities $\prob_k$ with $\prob_k\ge \eps/\threshold$.

\begin{figure}[ht]
\begin{LPTAB}
\begin{LP}{Alternative to $\MPG$}
\begin{array}{rr}
\text{\em if } i<\threshold: & x_i + w_i 
\\
\text{\em if } i\ge\threshold: & \max(x_i,w_i)
\end{array}
\Big\}
 & \le & r_i + \displaystyle\sum_{j<i} \sigma_{i-j}w_j & (i \in \limitedlevels) 
\\
 \sum_{k} y_{ki} & = & x_i & (i \in \limitedlevels)
\\
 \sum_{i} y_{ki} & = & 1 & (k\in[n])
\smallskip
\\
w_i, x_i, \,y_{ki} & \ge  & 0 & (i \in \limitedlevels, k\in[n])
\smallskip
\\  y_{ki} & \in  & \{0,1\} & (i \in [\threshold-1], k:\prob_k \ge \eps/\threshold)
\end{LP}
\end{LPTAB}
\caption{A practical alternative to $\MPG$ with integrality gap $1+O(\eps)$.}\label{fig:alternative}
\end{figure}

Solving this mixed program suffices,
because any near-optimal fractional solution $(x,w,y)$ to it
can be rounded to a near-optimal integer solution
(corresponding to a near-optimal $\threshold$-relaxed code):
\begin{lemma}
  Given any fractional solution $(w,x,y)$ 
  to the mixed program in Fig.~\ref{fig:alternative},
 one can compute 
 in $O(n)+O_\eps(\polylog n)$ time
 an integer solution 
 $(\hat w, \hat x, \hat y)$ of cost
 at most $1+O(\eps)$ times the cost of $(w,x,y)$.
\end{lemma}

{\em Proof sketch.}
For each $i<\threshold$, in increasing order, 
if $x_i$ and $\sum_k y_{ik}$
have fractional part $f>0$, do the following.
Let $i' = i+\cost(0)$.
Decrease $x_i$ by $f$,
increase $w_i$ by $f$,
and increase $x_{i'}$ by $f$.
(This preserves the capacity constraint because increasing $w_{i'}$ by $f$
increases the right-hand side of the capacity constraint for $i'$ by at least $f$,
since $\sigma_{i'-i} = \sigma_{\cost(0)} \ge 1$.)
Also, decrease $\sum_k y_{ki}$ by $f$
and increase $\sum_k y_{ki'}$ by $f$
by (repeatedly, if necessary) decreasing the (non-integral) $y_{ki}>0$ with smallest $\prob_k$,
and increasing the corresponding non-integral $y_{ki'}$.

Since these non-integral $y_{ki}$'s have $\prob_k < \threshold/\eps$,
for each $i$, the increase in the cost is at most $ (\eps/\threshold)f \cost(0)$, 
which is less than $(\eps/\threshold)\cost(0)$,
so the total increase in the cost (for all levels $i<\threshold$) is at most $\eps\cost(0)$.

After this modification, each $x_i$ for $i<\threshold$ is an integer.
Take $(\hat w, \hat x,\hat y)$ to be an optimal, all-integer greedy extension of this assignment 
to these $x_i$'s.
That is, for each $i\in\limitedlevels$, in increasing order,
take $\hat w_i$ maximally subject to the capacity constraint,
and, if $i\ge\threshold$, take $\hat x_i$ maximally subject to the capacity constraint.
Then take $\hat y$ so that the corresponding code is monotone.
This greedy extension is optimal by an argument similar to the proof of Lemma~\ref{lemma:hulg extension},
so it has cost at most the cost of the modified $(w, x,y)$.
\endproof

\paragraph{Finding a $(1+\eps)$-approximation is in NC}
Given that \HUL is neither known to be in P (polynomial time), 
nor known to be NP-hard,
it is interesting that the results in this paper extend
to show that, given any fixed $\eps$, the problem of $(1+\eps)$-approximating \HUL is in NC
(Nick's class --- polynomially many parallel processors
and polylogarithmic time).
 (For instances in which $\cost(1)\le 3/\eps$,
the cumulative distribution $\PROB$ 
and the signatures $r$ and $\sigma$
necessary for Thm.~\ref{thm:cost1sig}
can be computed in NC,
and the remaining computation takes time $O_\eps(\polylog n)$
on one processor.
For instances with no restrictions on the cost,
one can use the fact that $\limitedlevels=O_\eps(\log n)$
to show that each $O(n)$-time step in the proof of Thm.~\ref{thm:main}
is in NC.)

\paragraph{Open problems}
The PTAS in this paper is not a fully polynomial-time approximation scheme (FPTAS).
That is, the running time is not polynomial in $1/\eps$.
Is there an FPTAS?
For that matter, 
is there a polynomial-time exact algorithm?
And, of course, is \HUL NP-complete?

\section*{Acknowledgements}
The authors are very grateful to the two anonymous referees for their patience and helpful comments.

\bibliographystyle{plain}
\bibliography{journal}

\section*{Appendix}~

{\bf Proof of Lemma~\ref{lemma:hulg solution to code}.}
  \noindent{\em Part (i).}
  Let $\forest$ be the forest in the tree representation of $\code$.  Each
  non-empty level $i$ in $\forest$ is in $\limitedlevels$, by
  Observation~\ref{obs:limitedlevels}.

  Let $x_i$ and $w_i$ be, respectively, the number of codewords and
  interior nodes in level $i$ of $\forest$.  Let $y$ be the assignment of
  codewords (or rather codeword costs) to probabilities: that is,
  $y_{ki} = 1$ iff $\cost(\code_k)=i$ (else $y_{ki}=0$).  Let $z$ be
  the assignment of levels to groups: that is, $z_{gi} = y_{ki}$ for
  all $i <\threshold$, $g\in[\gamma]$, and $k\in \group_g$.

  First consider the capacity constraint of $\MPG$.  Level $i$ of $\forest$
  has at least $x_i+w_i$ nodes, or $\max(x_i, w_i)$ if $i\ge
  \threshold$.  Up to $r_i$ of these nodes can be parentless in $\forest$
  because they are roots in $\universe$.  Each of the rest has a
  parent in $\forest$ that is an interior node in $\forest$ in a level $j<i$.
  There are at most $\sum_{j<i} \sigma_{j-i} w_j$ nodes with such
  parents, because each of the $w_j$ interior nodes in a given level
  $j$ of $\forest$ can parent at most $\sigma_{i-j}$ nodes in level $i$ (one
  for each of the $\sigma_{i-j}$ letters of cost $i-j$ in
  $\alphabet$).  Thus, the capacity constraint is met.  By inspection,
  $(w,x,y,z)$ meets the remaining constraints of $\MPG$, and the cost
  of $(w,x,y,z)$ is $\cost(\code)$.  This proves Part (i) of the
  lemma.

  \smallskip
  \noindent{\em Part (ii).}
  Given any set $\code\subset\universe$, let $\code_{<\threshold}$
  denote $\{\code_k ~|~\cost(\code_k) < \threshold\}$.

  Start with $\code\leftarrow \emptyset$.  For each $i \in
  \limitedlevels$, in increasing order, add to $\code$ any $x_i$
  strings from level $i$ of $\universe$ that have no prefix in
  $\code_{<\threshold}$.

  This construction clearly generates a $\threshold$-relaxed code as
  long as there are enough strings available to assign in each level.
  There will be, because the construction maintains the following
  invariant: {\em for each $j<i$, at least $w_j$ strings in level $j$
    of $\universe$ are available.}  (Recall that a string is available
  if it has no prefix in $\code_{<\threshold}$.)  Suppose this
  invariant holds before codewords are added from level $i$.  At that
  point, the number of available strings in level $i$ of $\universe$
  must be at least the right-hand side of the capacity constraint for
  $i$.  In the case that $i < \threshold$, since the capacity
  constraint holds, the right-hand side is at least $x_i+w_i$, so
  placing $x_i$ of the available strings into $\code$ leaves $w_i$
  still available, maintaining the invariant.  In the case that $i \ge
  \threshold$, the right-hand side is both at least $w_i$ (so the
  invariant is maintained) and at least $x_i$ (so there are $x_i$
  available strings to add to $\code$, without making any string
  unavailable, since $i \ge \threshold$).

  Finally, assign to each probability $\prob_k$ a codeword from $\code$ of
  cost $i'$ such that $y_{ki'} = 1$.  (This is possible because in
  $\code$ there are $x_i = \sum_{k} y_{ki}$ codewords of each cost
  $i$.)  Then, $\cost(\code)$ equals the cost of $(w, x, y, z)$.
\endproof

\medskip
{\bf Proof of Lemma~\ref{lemma:hulg extension}.}
Let $(w,x,y,z)$ be any minimum-cost feasible extension of $z$.
Let $(\hat w, \hat x, \hat y, z)$ be the greedy extension (if it is well defined).

Given $z$, the constraints of $\MPG$ force $\hat x_i = x_i$ for $i<\threshold$.

By induction on $i \in \limitedlevels$
(using the maximality of $\hat w_i$
and that $\hat x_i = x_i$ for $i<\threshold$),
it follows that $\hat w_i\ge w_i$ for all $i\in\limitedlevels$.
Thus, replacing $w$ by $\hat w$ in $(w,x,y,z)$ gives
a solution $(\hat w, x, y, z)$ that is also feasible and optimal.

Now suppose for contradiction that $x_{i} \ne \hat x_{i}$ for some level $i$.
Fix $i'$ to be the minimum such level.
Note that $i'\ge \threshold$ since $\hat x_i = x_i$ for $i<\threshold$.
Since $\hat x_i = x_i$ for $i<i'$,
and $\hat x_{i'}$ is maximal (by definition of the greedy extension),
it follows that $x_{i'} < \hat x_{i'}$.
Thus,
the capacity constraint for level $i'$ ($\ge\threshold$) is loose for $(\hat w,x,y,z)$.
Increasing $x_{i'}$ by 1, and decreasing $x_{j}$ by 1 for some $j>i$ 
(and adjusting $y$ accordingly) gives a feasible solution 
that is cheaper than $(\hat w, x, y, z)$,
contradicting the optimality of $(\hat w,x,y,z)$.

Thus, $\hat x = x$.  Thus, $(\hat w, \hat x, y, z)$ is feasible.  By the choice of $\hat y$
in the definition of the greedy extension, the lemma follows.
%
%
\endproof

\end{document}